# Impact of Reactor Neutron Spectrum on Measured Spectrum Averaged Cross Sections

Michal Kostal[1], Evzen Losa[1,2], Stanislav Simakov[3], Martin Schulc[1], Jan Simon[1], Vojtech Rypar[1], Martin Mareček[1], Jan Uhlíř[1], Tomáš Czakoj[1], Andrej Trkov[4], Roberto Capote[5]

[1] Research Center Rez, 250 68 Husinec-Rez 130, Czech Republic
[2] Dept. of Nuclear Reactors, Faculty of Nuclear Sciences and Physical Engineering, Czech Technical University in Prague, V Holesovickach 2, Prague 180 00, Czech Republic
[3] Institute for Neutron Physics and Reactor Technology, Karlsruhe Institute of Technology, Hermann-von-Helmholtz-Platz 1, D-76344 Eggenstein-Leopoldshafen, Germany
[4] Jožef Stefan Institute, Jamova cesta 39, 1000 Ljubljana, Slovenia
[5] Nuclear Data Section, International Atomic Energy Agency, A-1400 Wien, Austria

Email: Michal.Kostal@cvrez.cz
Telephone: +420266172655

Abstract
The cross section averaged over $^{235}$U thermal-neutron induced fission spectrum is a fundamental quantity that can be used in evaluation and validation of nuclear data. Many experiments focused on the determination of Spectrum Averaged Cross Sections (SACS) in $^{235}$U($n_{th}$,f) Prompt Fission Neutron Spectrum (PFNS) in light water reactors using enriched uranium fuel. In these reactors, there is already some amount of water moderator between the uranium fuel and the irradiated sample. Due to the decrease of hydrogen cross-section with neutron energy, the high energy tail of the reactor spectrum in cores with water moderator may be harder than the pure PFNS. This paper aims to compare the shape of the actual reactor spectrum in various core positions of a research light-water reactor differing each from other by the effective water thickness. The spectrum shape is determined both by calculations and experimentally using various high energy threshold reactions.
The impact of the photo-nuclear reactions (γ,n) competing with (n,2n) in production of the same residual nucleus was shown to be less than a percent for most of studied dosimeters. An important exception was found for $^{197}$Au(n,2n) $^{196}$Au dosimeter irradiated in the outcore channel where a notable photo-neutron contribution to the production of $^{196}$Au is caused by the neutron production from the high energy γ-rays from thermal-neutron capture in $^{54}$Fe. The corresponding ENDF/B-VIII.0 data turned out to underestimate such γ-yield by 40% in comparison with ENDF/B-VI.8. This has improved but however not resolved the disagreement between our measurement and calculations. The remaining deficiency was attributed to the underestimation of the evaluated cross section IAEA/PD-2019 for the $^{197}$Au(γ,n) cross section near the reaction threshold. The later was confirmed by comparison with existing measured data.

## 1 Introduction

A large set of SACS measurements in light water reactors was performed in past (Abd 2010, Arribére, et al. 2001, Bruggeman et al. 1974, WÖLFLE et al 1980, Firestone et al. 2017, Dorval et al 2006 or Maidana et al 1994). It is assumed within these experiments that the reactor spectrum above 2 MeV Steinnes 1970 or 2.6 MeV Suarez et al 1997 is almost identical with $^{235}U$ thermal-neutron induced PFNS (Prompt Fission Neutron Spectrum). The experiments performed in the LR-0 reactor show that reactor spectrum in the reactor using VVER-1000 fuel is undistinguishable from $^{235}U(n_{th}$, fiss.) PFNS in the region above 6 MeV Kostal et al 2018. However, the reaction-rate ratios of reactions with lower energy threshold $^{58}Ni(n,p)$ and $^{27}Al(n,\alpha)$ in the LR-0 are still very close to the ratio obtained in $^{235}U$ PFNS. This observation was generally confirmed by recent experiments performed in the VR-1 reactor Kostal et al 2021, but in higher energy regions, it was found that the actual fast part of the VR-1 reactor spectrum is harder than $^{235}U$ PFNS.

Different observations made at the experimental VVER-1000 fueled reactor LR-0 are caused by the differences in the fuel structure and uranium enrichments. VR-1 reactor uses tubular fuel with enrichment near 20 wt.% of $^{235}U$ and fuel tubes based on $UO_2$ dispersed in the aluminum matrix, whereas LR-0 core contains classical pin-type fuel with LEU (3 – 4.4 wt.% of $^{235}U$) $UO_2$ pellets and cladding from zirconium alloy. The comparison of actual reactor spectra in the central core position of VR-1 and LR-0 and their difference from the PFNS is plotted below in Figure 1.

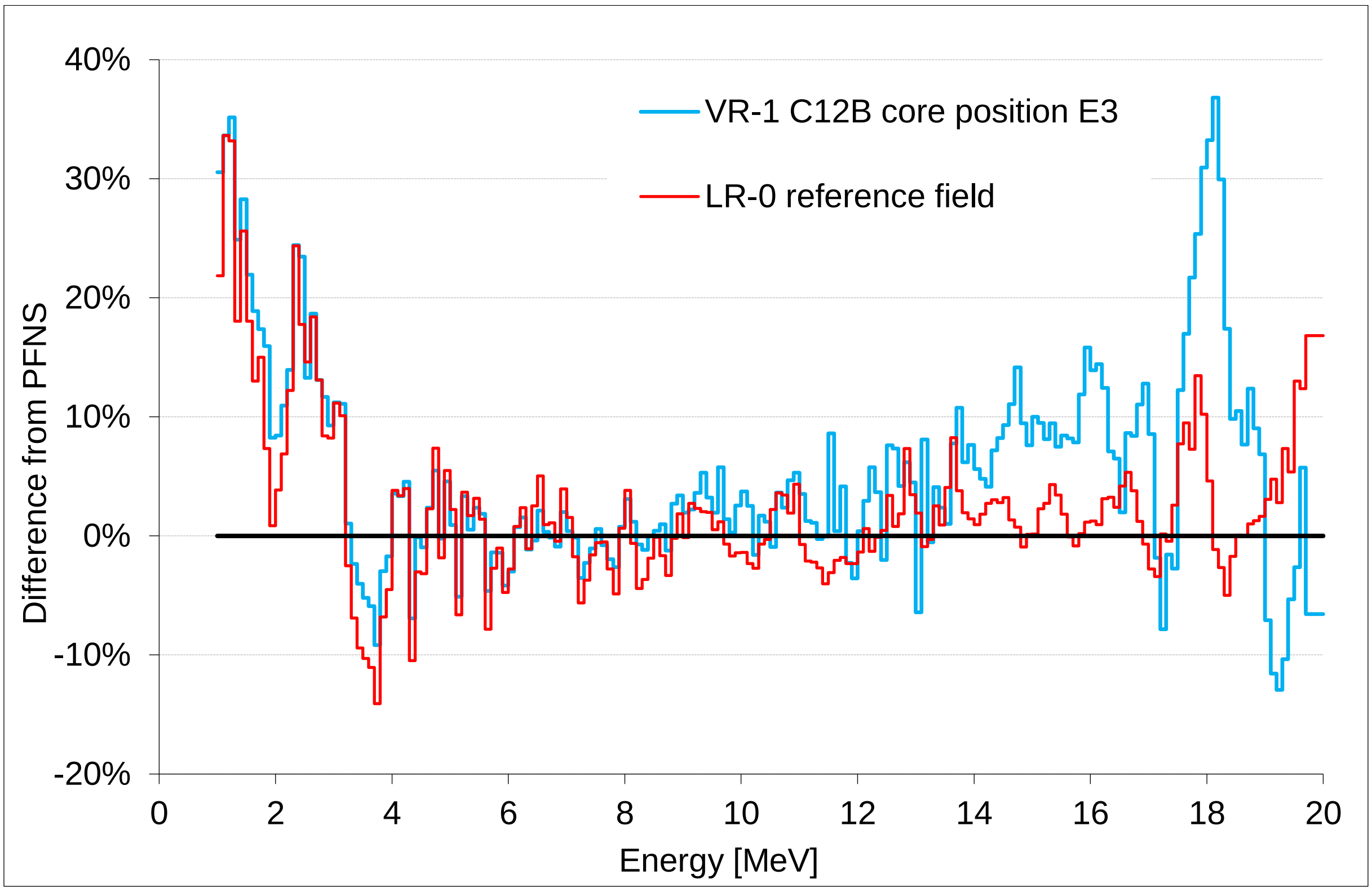

Figure 1: Comparison of VR-1 and LR-0 reactor spectra with $^{235}U$ PFNS.

The differences in the high-energy tail of the fission spectra in the LR-0 and VR-1 reactors compared to the pure $^{235}U$ prompt fission neutron spectrum can be explained by the differences in the macroscopic cross sections of the homogenized cores (see Figure 2). The total cross section of the homogenized LR-0 core is higher compared to VR-1. Also, the VR-1 homogenized cross section has decreasing character in the region above 15 MeV. The combination of both facts causes a non-negligible decrease in interaction rates with increasing energy in the VR-1. The situation in LR-0 is different because the cross section is nearly constant, and the oscillations from the average cross section are relatively lower (in percentage terms); thus, the interaction rate is nearly constant. This fact is propagated in nearly constant energy-dependent attenuation of high energy neutrons (> 10 MeV) in the LR-0, while in the VR-1, it decreases. Reflection of this fact is a harder tail of spectrum above 15 MeV in the VR-1 reactor.

Further text concerns with the measurement of reaction rates and following evaluation of the SACS in core locations of the VR-1 reactor with various effective water layer thicknesses. Different core positions were selected for study: fully surrounded by fuel, positions in the boundary fuel assemblies, and positions with water gap between fuel and irradiation position. Experimental data were then compared with simulations. It was found out that in the positions fully surrounded by fuel, the spectrum has a similar shape as in the previous experiment presented in Kostal et al 2021. Therefore, the current results for the central core position can serve as a verification of previous experiments.

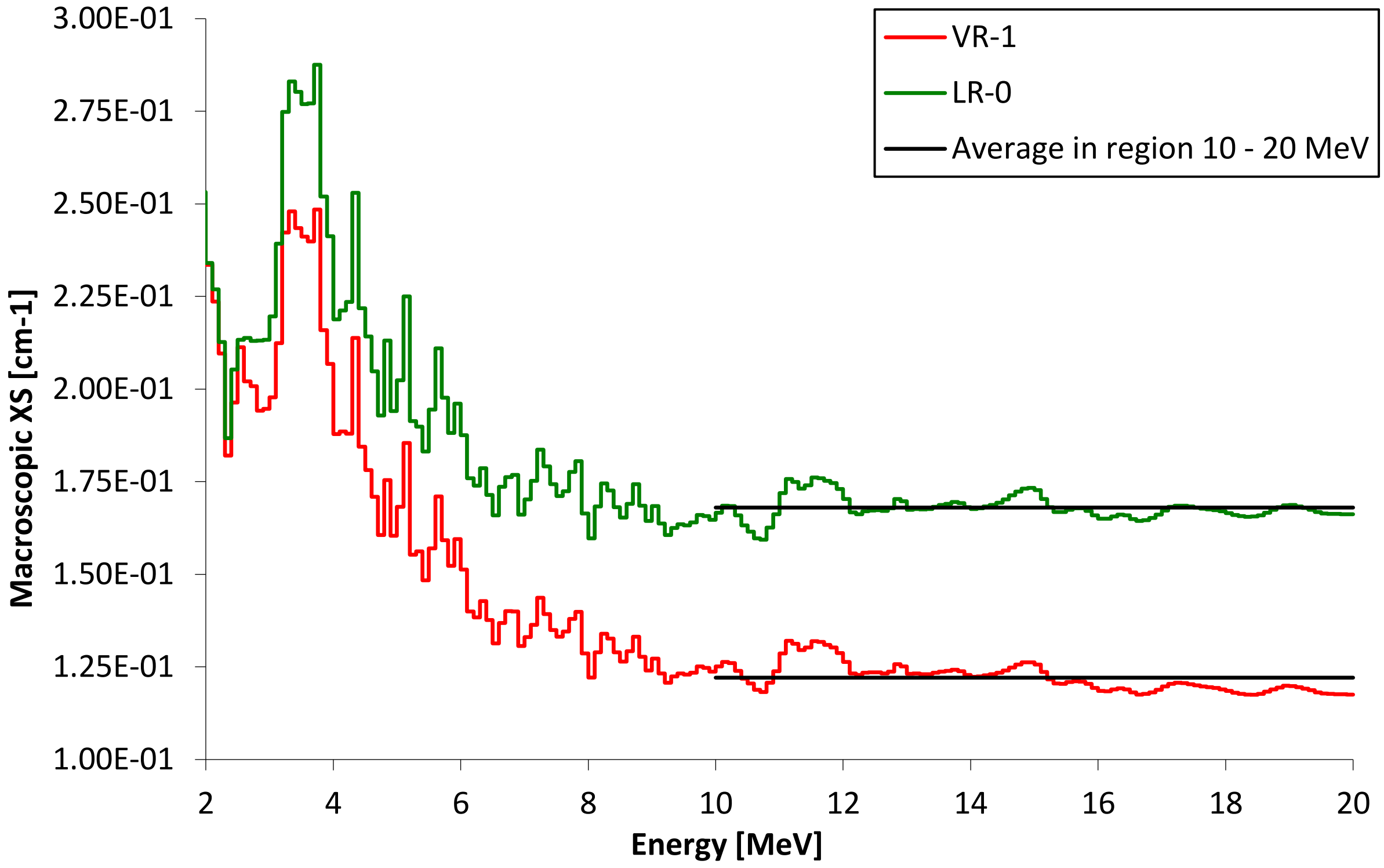

Figure 2: Macroscopic cross sections of VR-1 and LR-0 cores

## 2 VR-1 reactor

The VR-1 research reactor is a light-water, zero-power pool-type reactor operated by the Czech Technical University in Prague. The core is composed of IRT-4M fuel enriched to 19.75 wt. % of $^{235}$U. Higher enrichment in combination with good moderation allows formation of the compact core, and the achievable fluxes are relatively high, considering the fact that the VR-1 is a zero-power reactor. The design of the reactor allows a maximum thermal power of approximately 650 W, which corresponds to a fast neutron flux above 1 MeV in the order of $1.4 \cdot 10^9$ $n \cdot cm^{-2} \cdot s^{-1}$ in the central region of the core.

The core is composed of fuel assemblies (IRT-4M type), dummy assemblies, and special stainless-steel assemblies (Frybort 2020, Czakoj 2021) formed by steel rods at the boundary of the reactor core. Several dry vertical channels with different diameters up to 56 mm are placed in the fuel and dummy assemblies. A horizontal radial channel is adjacent to the reactor core but was not utilized for current experiments. To study the effect of water on the high energy tail of $^{235}$U PFNS, dosimeters sets using various reactions with different sensitivities (Figure 8) were placed at different core positions with different effective water thicknesses. Materials of the irradiated targets include $CF_2$ and Au to study the (n,2n) reactions, Al and Ni foils for power monitoring and $^{58}$Ni(n,2n) and $^{58}$Ni(n,x)$^{57}$Co studies, and further Mg, Ti, and Fe for studies of (n,p) reactions. In total, six sets of foils have been used in different positions of the reactor core. All the sets were always centered axially to be in the maximum flux position.

The radial positions for foil-stack irradiation were chosen to correspond in pairs to a very similar shape of the neutron spectrum and are given in Table 1. These are the two positions in the centre of the core (pos.2 and 4 – E3 and E5 in the figure 3), the two positions at the boundary (pos.1 and 3 – B3 and G3 in the figure 3) and the lateral position 5 (A5 in the figure 3) behind the water layer (water layer ~ 7 cm) and position 6 in the upper corner (B8 in the figure 3), one row behind the fuel (water layer ~ 10 cm). Between the foil positions, there are only small differences in the position of surrounding fuel assemblies, dry channel diameters, or water layer thicknesses, as seen in Figure 3.

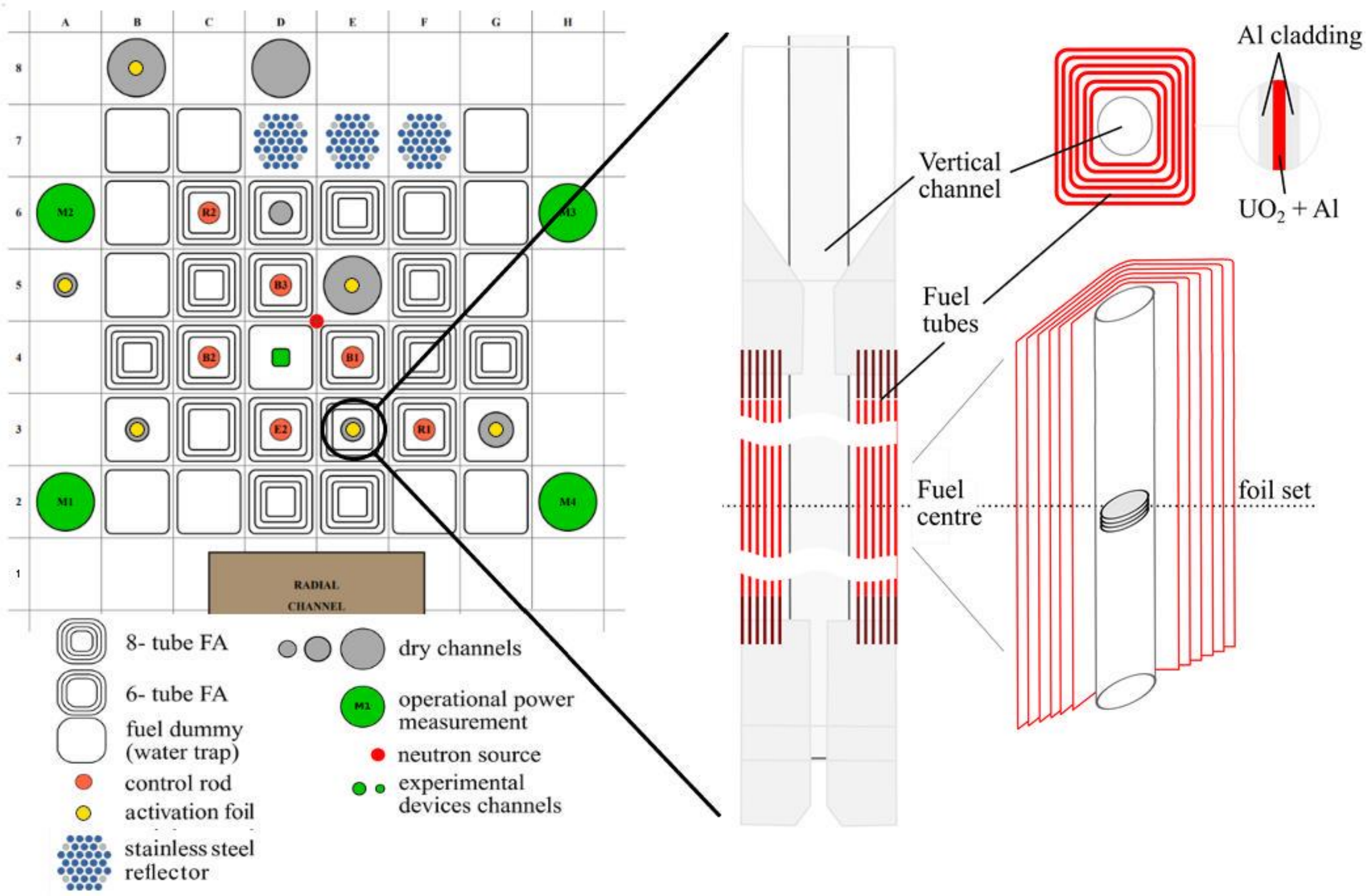

Figure 3: View on VR-1 reactor core (C16) used in the first experiment, together with scheme of target arrangements, see the Table 1 for foil numbers and core positions.

Table 1.: Positions of the foil sets in the C16 core

| Set | Position in the core | Description of the location | Materials in the set |
|---|---|---|---|
| 1 | B3 | Core boundary | $CF_2$, Ni, Au, Mg, Ti, Fe, Al |
| 2 | E3 | Surrounded by fuel | |
| 3 | G3 | Core boundary | |
| 4 | E5 | Surrounded by fuel | |
| 5 | A5 | In water reflector | |
| 6 | B8 | In water reflector | |

## 3 Experimental and calculation methods

### 3.1 Irradiation setup

The neutron spectrum at six irradiation positions was studied using well-defined dosimetry reactions from the IRDFF library (Trkov et al 2020), which were also previously validated in the VR-1 reactor (Kostal et al 2021). Each set of the activation foils always formed a stack of successively alternating materials – flux monitors and those for the study of the dosimetry reaction. The stacks were identical in each position to minimize possible measurement uncertainties; very thin Al and Ni monitors, using extremely well-known $^{27}$Al(n,α) and $^{58}$Ni(n,p) reactions, were positioned between the dosimetry foils for normalization and for testing the homogeneity of neutron flux in the target. The scheme of the activation foil stack, which forms the target for neutron irradiation, is plotted in Figure 4. The details of the spectra calculations and their specific features are reported in Section 3.3.

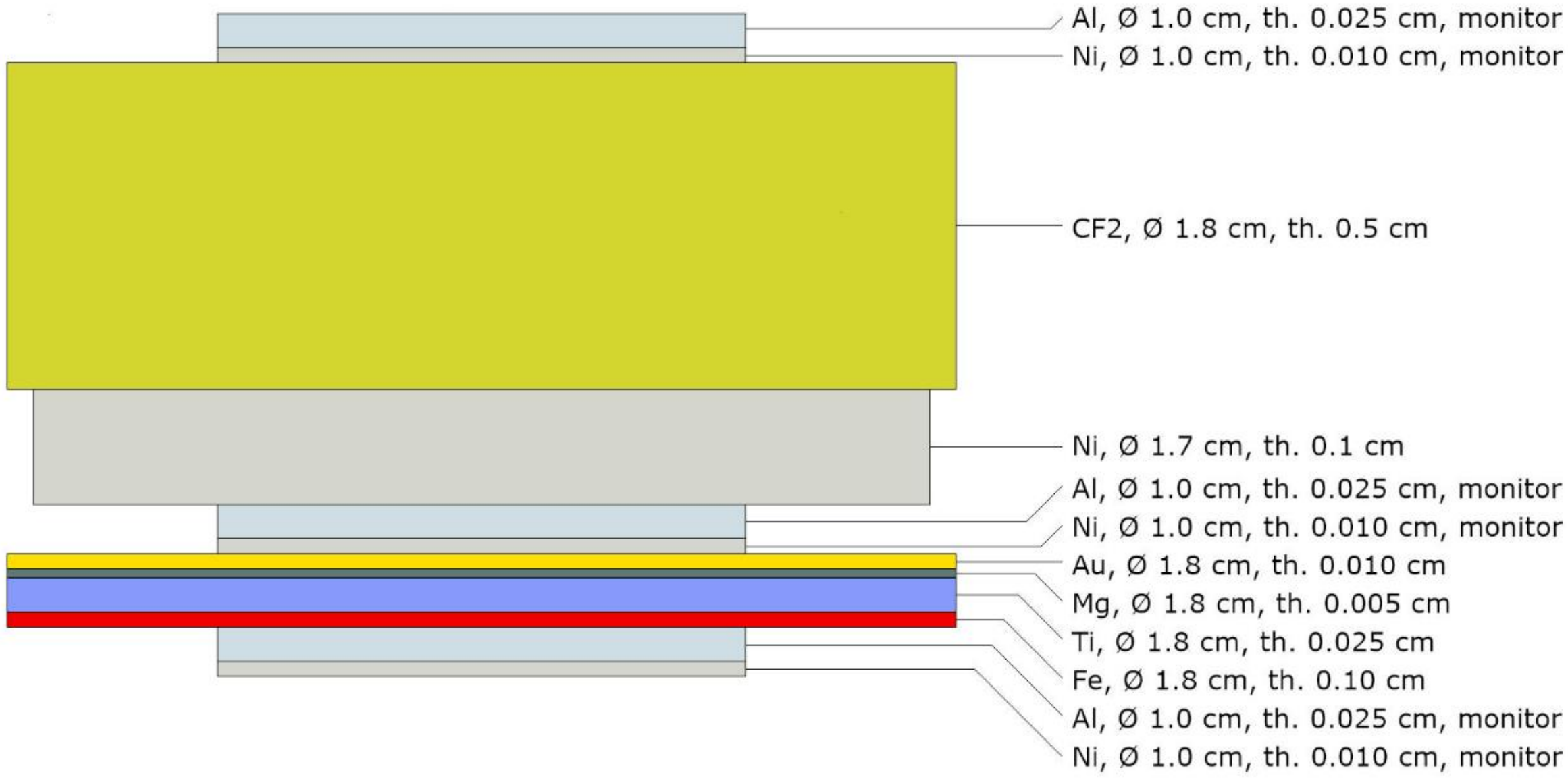

Figure 4: Foils forming target assembly stack.

Because the power and related neutron flux are sufficient to activate the threshold reactions, small activation foils placed in the stack can be used. Thin Ni (D = 1 cm, th.= 0.1 mm) and Al (D = 1 cm, th.= 0.25 mm) were placed between the activation foils for SACS measurements. The thickness of $CF_2$ and the adjacent Ni monitors was 5 mm and 1 mm, respectively. Inhomogeneities in the neutron field could therefore impact the evaluated reaction rates and monitors were used for the evaluation of such effect.

The flux was derived using average monitors reaction rate by the approach described in Kostal et al 2021. In the centrally located stacks the flux above 1 MeV was ~ $7\cdot10^9$ $n\cdot cm^{-2}\cdot s^{-1}$ and the flux above 10 MeV ~ $1.3\cdot10^7$ $n\cdot cm^{-2}\cdot s^{-1}$. Because each target contains more monitoring foils, due to the use of thick dosimeters, the comparison of the actual reaction rate with the average of all monitors in the stack was used to test the flux homogeneity in the target, see Table 2. The so estimated flux variations are negligible in the selected positions relative to the core center. This allows to assume that the flux is distributed homogenously in the whole target.

Table 2.: Flux homogeneity between variously positioned monitors in used targets (differences from the average)

| | $^{27}Al(n,\alpha)$ | | | | | |
|---|---|---|---|---|---|---|
| | Set 1 | Set 2 | Set 3 | Set 4 | Set 5 | Set 6 |
| Upper | 2.6% | -0.8% | -0.1% | -0.5% | -0.7% | 0.4% |
| In center | -3.4% | 1.1% | -2.5% | -0.5% | -3.2% | -1.4% |
| Lower | 0.8% | -0.4% | 2.6% | 1.1% | 3.9% | 1.0% |
| | $^{58}Ni(n,p)$ | | | | | |
| | Set 1 | Set 2 | Set 3 | Set 4 | Set 5 | Set 6 |
| Upper | 1.8% | 0.7% | -0.8% | -1.1% | 1.4% | -0.4% |
| In center | -0.8% | -0.1% | -1.5% | 0.4% | -1.5% | -0.7% |
| Lower | -1.0% | -0.7% | 2.3% | 0.7% | 0.1% | 1.1% |

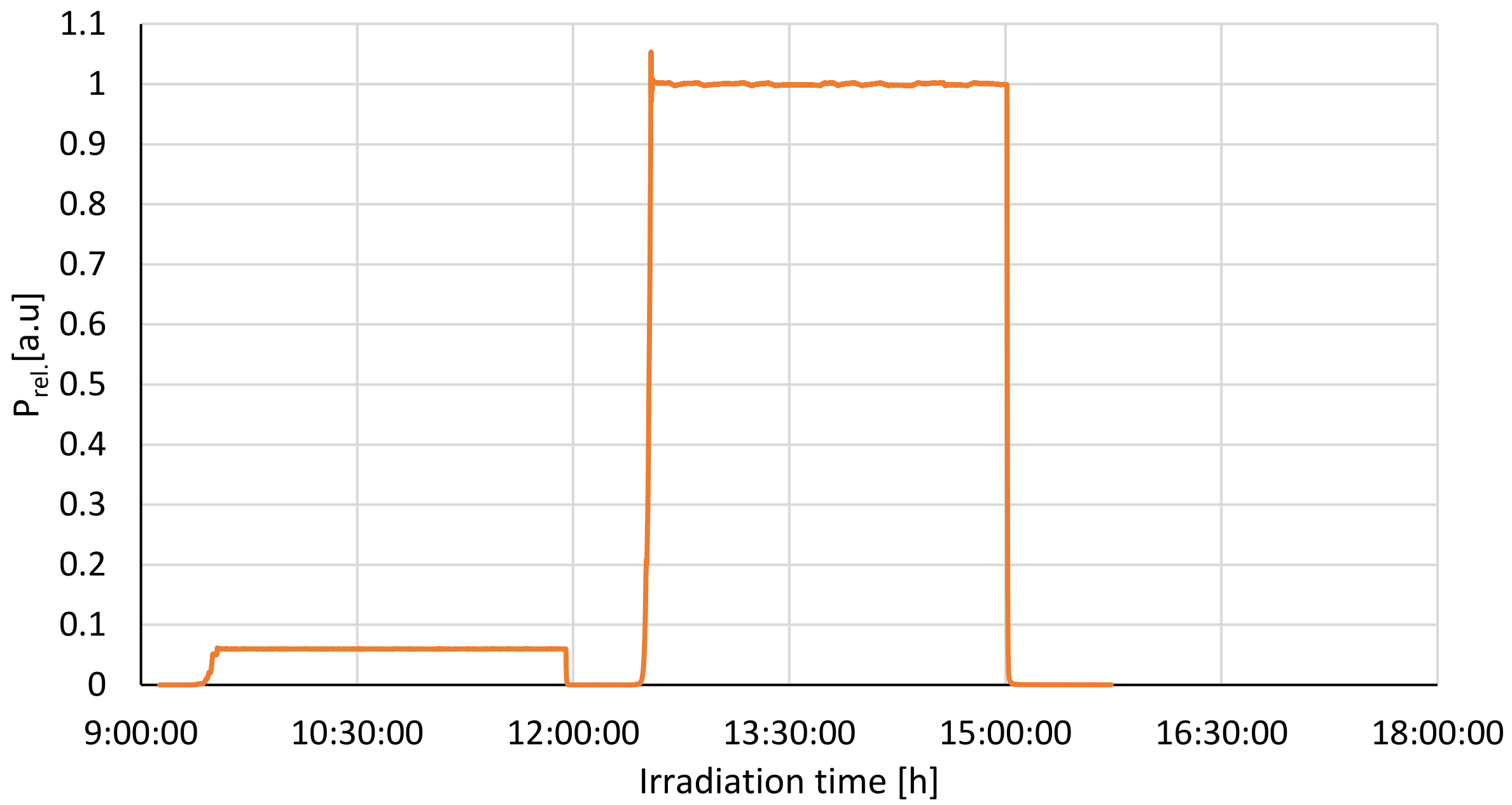

Figure 5: Power profile during irradiation

The power profile during foils irradiation was not constant, as visible in Figure 5, but the differences from the average power during steady state operation are small, of order of 1 %. Due to the piecewise continuous and nearly constant character of power evolution during the experiment, the precise $A/A_{sat}$ was used, as defined by equation (1).

$$\frac{A(\overline{P})}{A_{Sat}(\overline{P})} = \sum_i \frac{P_i}{\overline{P}} \times \left(1 - e^{-\lambda.T_i}\right) \times e^{-\lambda.T_i^{End}} \qquad (1)$$

Here $A/A_{sat}$ is the ratio of the activity to saturated activity, $\frac{P_i}{\overline{P}}$ is the relative power in i-th interval of the irradiation period, $T_i$ is the irradiation time in i-th interval of the irradiation period and $T_i^{End}$ is the time from the end of the i-th irradiation interval to the end of irradiation period.

## 3.2 Gamma spectrometry

Experimental reaction rates were determined from the measured Net Peaks Areas (NPA), determined $A/A_{sat}$ ratio (see Eq. 1), calculated detector efficiency, and tabulated constants, see Eq.(2). The NPAs were measured employing HPGe gamma spectrometry. Dosimetry foils, as well as monitoring foils, were measured separately with a well-characterized HPGe detector in the Research Center Rez (see Kostal et al 2018b). The thinner foils (Mg, Fe, Ti, Au, Ni) and Al monitors were fixed in the plastic EG-3 type holder for ensuring the measuring geometry. The thicker Ni and $CF_2$ foils were placed on the top of a coaxial HPGe detector cap, and in the Ni case also 2 cm above the cap. To suppress the background signal, the detector was placed in a lead shielding with a thin inner copper lining and rubber coating. Despite the shielding, the background without any sample was measured and subsequently subtracted from the sample spectrum.

Genie 2000 software (Canberra) was used for the spectrum evaluation. The efficiency curve and appropriate Coincidence Summing Factors (Tomarchio et al 2009 ) were determined using the MCNP6.2 code and a validated mathematical model of HPGe detector according to the methodology previously established and validated in Dryak et al 2006 and Boson et al 2008. The input parameters used in evaluations are listed in Table 3. The detector efficiency uncertainty was assessed from the difference between the experimentally determined efficiency and the efficiency determined with the precise HPGe model and is about 1.9 % in the relevant energy interval (Kostal et al 2018b).

A set of Monte Carlo simulations was performed for the point calibration source represented by EG3 holder. The resulting efficiency curve for the EG3 type source is plotted in Figure 6 and the differences from the actual efficiency are listed in Table 4. It can be said that the efficiencies of thin foils in the EG3 holder are comparable to those ones for the calibration source and the use of detector efficiency from the calculation is justified. The most notable efficiency differences are for thicker foils (Ni 122.1 keV peak and $CF_2$ 511 keV peak). It reflects the significant attenuation of the lower-energy gamma in the first case and the higher average distance from the detector in the second case.

The detector energy calibration was performed before the experiment using the standard point sources $^{60}$Co, $^{88}$Y, $^{133}$B, $^{137}$Cs, $^{152}$Eu, and $^{241}$Am with an uncertainty less than 1.0 keV throughout the used energy range.

$$q(\overline{P}) = \left(\frac{A(\overline{P})}{A_{\mathrm{Sat}}(\overline{P})}\right)^{-1} \times \frac{NPA}{T_{Live}} \times \frac{1}{\varepsilon \times \eta \times N} \times \frac{\lambda \times T_{real}}{(1 - e^{-\lambda . T_{real}})} \times \frac{1}{e^{-\lambda . \Delta T}} \times \frac{1}{CSCF} \qquad (2)$$

Where:

$q(\overline{\mathrm{P}})$; is the reaction rate of activation during power density $\overline{\mathrm{P}}$ (power in first day of irradiation experiment);

NPA; Net Peak Area

$\mathrm{T}_{Live}$; is time of measurement by HPGe, corrected to detector dead time;

$\mathrm{T}_{\mathrm{real}}$; is time of measurement by HPGe, corrected to detector dead time;

$\Delta\mathrm{T}$ ; is the time between the end of irradiation and the start of HPGe measurement;

$\lambda$; is decay constant of studied isotope;

$\varepsilon$ ; is the gamma branching ratio;

$\eta$ ; is the detector efficiency (the result of MCNP6 calculation);

$N$ ; is the number of target isotope nuclei;

CSCF; Coincidence Summation Correction Factor

Table 3.: Summary of used monitors and detectors.

| Reaction | Peak [keV] | Material | Dimensions [mm] | Geometry | Detection Efficiency | CSCF | A/Asat. |
|---|---|---|---|---|---|---|---|
| $^{27}Al(n,\alpha)^{24}Na$ | 1368.6 | Al | D=10, th.0.25 | EG-3 on cap | 2.951E-2 | 0.863 | 1.144E-1 |
| $^{58}Ni(n,p)^{58}Co$ | 810.8 | Ni | D=10, th. 0.1 | EG-3 on cap | 4.549E-2 | 0.937 | 1.076E-3 |
| $^{56}Fe(n,p)^{56}Mn$ | 846.8 | | | | 4.311E-2 | 0.939 | 5.013E-1 |
| | 1810.7 | Fe | D=1.8, th 0.1 | EG-3 on cap | 2.293E-2 | 0.816 | 5.013E-1 |
| $^{54}Fe(n,p)^{54}Mn$ | 834.8 | | | | 4.375E-2 | 1.000 | 2.442E-4 |
| $^{24}Mg(n,p)^{24}Na$ | 1368.6 | Mg | D=1.8, th 0.05 | EG-3 on cap | 2.920E-2 | 0.864 | 1.144E-1 |
| $^{19}F(n,2n)^{18}F$ | 511.0 | $CF_2$ | D=18, th 5 | On cap | 6.071E-2 | 1.000 | 6.230E-1 |
| $^{197}Au(n,2n)^{196}Au$ | 333.0 | Au | D=15, th. 0.1 | EG-3 on cap | 9.450E-2 | 0.809 | 1.228E-2 |
| | 355.7 | | | | 8.943E-2 | 0.987 | 1.228E-2 |
| $^{47}Ti(n,p)^{47}Sc$ | 159.4 | | | | 1.636E-1 | 1.000 | 2.249E-2 |
| $^{46}Ti(n,p)^{46}Sc$ | 889.3 | Ti | 10×10×0.25 | EG-3 on cap | 4.130E-2 | 0.830 | 9.096E-4 |
| $^{48}Ti(n,p)^{48}Sc$ | 983.5 | | | | 3.805E-2 | 0.668 | 4.098E-2 |
| | 1037.5 | | | | 3.644E-2 | 0.661 | 4.098E-2 |
| $^{58}Ni(n,x)^{57}Co$ | 122.1 | | | | 5.062E-2 | 1.000 | 2.806E-4 |
| $^{58}Ni(n,p)^{58}Co$ | 810.8 | Ni | D=17, th. 1 | 2.56 cm above cap | 1.453E-2 | 0.978 | 1.076E-3 |
| $^{58}Ni(n,2n)^{57}Ni$ | 1377.6 | | | | 9.550E-3 | 0.927 | 5.002E-2 |
| | 1919.5 | | | | 7.201E-3 | 0.939 | 5.002E-2 |
| $^{58}Ni(n,x)^{57}Co$ | 122.1 | | | | 1.465E-1 | 1.000 | 2.806E-4 |
| $^{58}Ni(n,p)^{58}Co$ | 810.8 | Ni | D=17, th. 1 | On cap | 4.617E-2 | 0.933 | 1.076E-3 |
| $^{58}Ni(n,2n)^{57}Ni$ | 1377.6 | | | | 3.014E-2 | 0.774 | 5.002E-2 |
| | 1919.5 | | | | 2.262E-2 | 0.806 | 5.002E-2 |

Table 4.: Difference between point calibration source (EG3 type) and the actual geometry of the source

| Peak [keV] | Foil | Geometry | Difference in efficiency | Difference in CSCF |
|---|---|---|---|---|
| 1368.6 | Al, D=10, th. 0.25 | EG-3 on cap | 1.5% | -0.2% |
| 810.8 | Ni, D=10, th. 0.1 | EG-3 on cap | 0.9% | -0.1% |
| 846.8 | Fe, D=18, th 0.1 | EG-3 on cap | 2.8% | -0.1% |
| 1810.7 | | | 2.4% | -0.4% |
| 834.8 | | | 2.5% | - |
| 1368.6 | Mg, D=18, th 0.05 | EG-3 on cap | 2.6% | -0.3% |
| 511.0 | CF2, D=18, th 5 | On cap | 12.0% | - |
| 333.0 | Au, D=15, th. 0.1 | EG-3 on cap | 5.9% | -0.7% |
| 355.7 | | | 5.5% | 0.0% |
| 159.4 | Ti, 10 × 10 × 0.25 | EG-3 on cap | 3.6% | - |
| 889.3 | | | 3.1% | -0.4% |
| 983.5 | | | 3.0% | -1.0% |
| 1037.5 | | | 3.1% | -1.1% |
| 122.1 | Ni, D=17, th. 1 | 2.56 cm above cap | 11.6% | - |
| 810.8 | | | 0.1% | 0.1% |
| 1377.6 | | | -0.2% | -0.1% |
| 1919.5 | | | -0.4% | 0.2% |
| 122.1 | Ni, D=17, th. 1 | On cap | 14.2% | - |
| 810.8 | | | -0.5% | 0.4% |
| 1377.6 | | | -1.2% | 1.1% |
| 1919.5 | | | -1.5% | 1.4% |

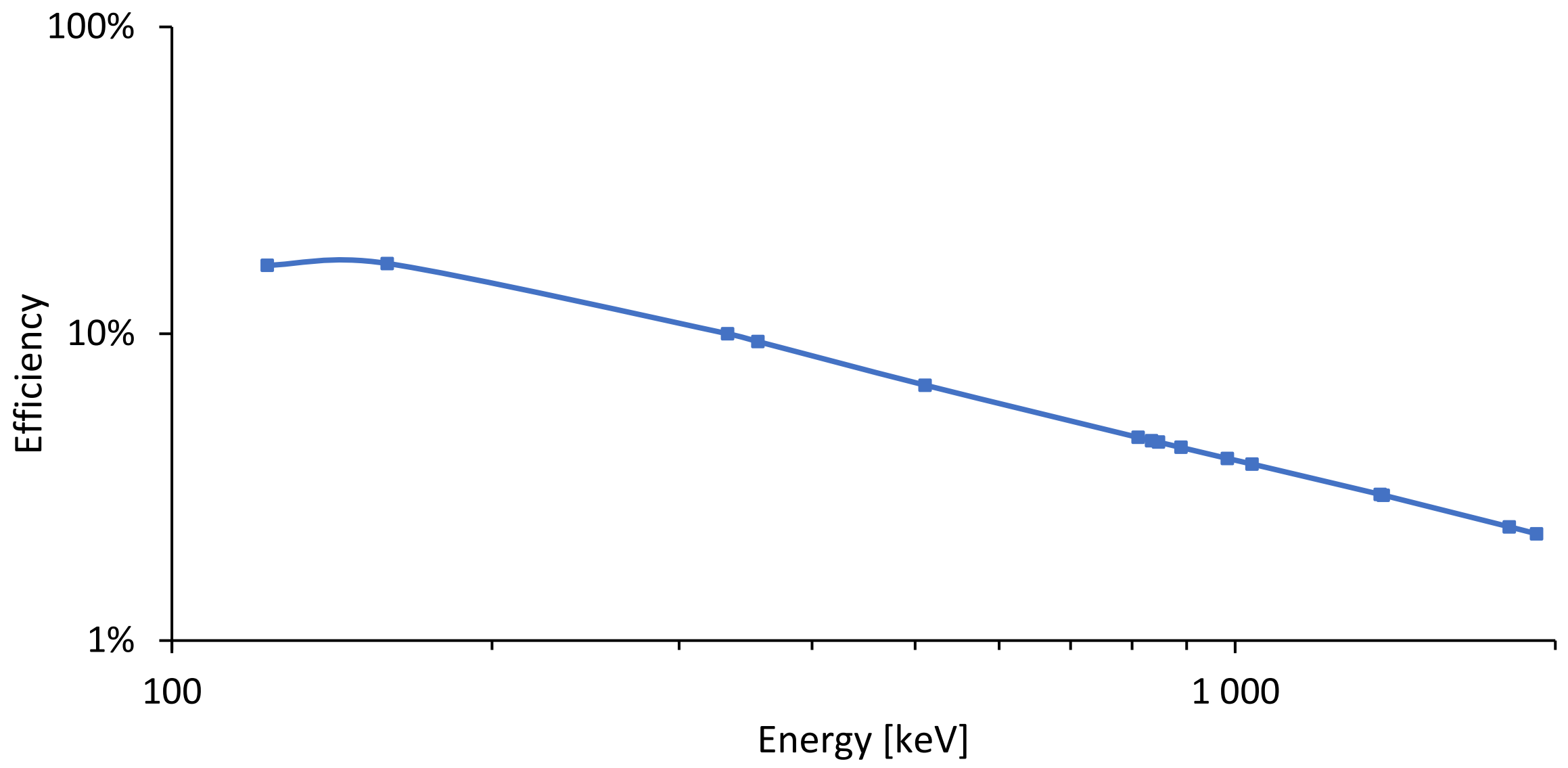

Figure 6: Calculated efficiency curve for point source of EG3 type on cap

The evaluated reaction rates are listed in Table 5. The related total uncertainties are listed in Table 6. The count rate uncertainty, being not higher than 1 %, includes the following main components: gross peak area, Compton continuum area, background area, and energy and peak shape calibrations parameters.

Besides above stated uncertainties, there are also other stochastic uncertainties: the radionuclide half-time value or branching ratios. However, these uncertainties are negligible in comparison with the count rate uncertainties.

Table 5.: Summary of measured reaction rates and monitoring reactions rates in reactor positions 1 - 6

| Position number | 1 | 2 | 3 | 4 | 5 | 6 |
|---|---|---|---|---|---|---|
| $^{58}Ni(n,2n)^{57}Ni$ | 1.34E-20 | 3.66E-20 | 1.35E-20 | 3.38E-20 | 5.96E-21 | 2.80E-21 |
| $^{19}F(n,2n)^{18}F$ | 2.49E-20 | 7.28E-20 | 2.57E-20 | 6.68E-20 | 1.11E-20 | 4.64E-21 |
| $^{27}Al(n,\alpha)^{24}Na$ | 2.02E-18 | 6.42E-18 | 2.10E-18 | 5.84E-18 | 8.41E-19 | 3.22E-19 |
| $^{24}Mg(n,p)^{24}Na$ | 4.37E-18 | 1.29E-17 | 4.45E-18 | 1.18E-17 | 1.87E-18 | 7.08E-19 |
| $^{48}Ti(n,p)^{48}Sc$ | 8.54E-19 | 2.74E-18 | 8.79E-19 | 2.47E-18 | 3.58E-19 | 1.33E-19 |
| $^{56}Fe(n,p)^{56}Mn$ | 3.12E-18 | 1.01E-17 | 3.25E-18 | 9.12E-18 | 1.29E-18 | 4.79E-19 |
| $^{46}Ti(n,p)^{46}Sc$ | 3.07E-17 | 9.99E-17 | 3.10E-17 | 8.92E-17 | 1.19E-17 | 4.16E-18 |
| $^{47}Ti(n,p)^{47}Sc$ | 4.79E-17 | 1.70E-16 | 5.08E-17 | 1.56E-16 | 1.73E-17 | 5.82E-18 |
| $^{197}Au(n,2n)^{196}Au$ | 1.47E-17 | 3.17E-17 | 1.35E-17 | 2.86E-17 | 6.94E-18 | 3.33E-18 |
| $^{58}Ni(n,p)^{58}Co$ | 2.80E-16 | 9.85E-16 | 2.96E-16 | 9.23E-16 | 1.04E-16 | 3.51E-17 |
| $^{58}Ni(n,x)^{57}Co$ | 7.77E-19 | 2.01E-18 | 7.41E-19 | 1.93E-18 | 3.23E-19 | 1.18E-19 |

Table 6.: Summary of uncertainties in measured reaction rates and monitoring reactions rates in reactor positions 1 - 6

| Position number | 1 | 2 | 3 | 4 | 5 | 6 |
|---|---|---|---|---|---|---|
| $^{58}$Ni(n,2n)$^{57}$Ni | 3.3% | 5.3% | 5.3% | 2.8% | 3.3% | 15.6% |
| $^{19}$F(n,2n)$^{18}$F | 1.9% | 1.9% | 2.0% | 2.0% | 2.0% | 2.1% |
| $^{27}$Al(n,α)$^{24}$Na | 3.1% | 2.0% | 2.8% | 2.0% | 3.4% | 2.1% |
| $^{24}$Mg(n,p)$^{24}$Na | 2.1% | 2.1% | 2.1% | 2.1% | 2.3% | 2.6% |
| $^{48}$Ti(n,p)$^{48}$Sc | 2.2% | 2.0% | 1.8% | 2.2% | 2.5% | 1.8% |
| $^{56}$Fe(n,p)$^{56}$Mn | 3.2% | 2.5% | 2.6% | 2.6% | 4.2% | 3.3% |
| $^{46}$Ti(n,p)$^{46}$Sc | 2.6% | 2.4% | 2.2% | 2.2% | 3.0% | 3.0% |
| $^{47}$Ti(n,p)$^{47}$Sc | 1.8% | 1.8% | 1.8% | 1.8% | 1.8% | 1.8% |
| $^{197}$Au(n,2n)$^{196}$Au | 6.8% | 3.7% | 5.4% | 4.1% | 5.9% | 6.8% |
| $^{58}$Ni(n,p)$^{58}$Co | 2.2% | 1.9% | 2.5% | 2.0% | 2.1% | 2.0% |
| $^{58}$Ni(n,x)$^{57}$Co | 5.7% | 6.8% | 5.0% | 5.8% | 6.9% | 15.2% |

### 3.3 Calculation methods and VR-1 spectra at irradiation point.

The MCNP6.2 Monte Carlo code (Werner et al 2017) in criticality mode was used for neutron and photon transport simulations. Resulting spectrum and reaction rates in the positions of activation foils have been obtained from the detailed model of the VR-1 reactor utilizing the ENDF/B-VIII.0 (Brown et al 2018) as the basic nuclear data library for the neutron transport. For reaction-rate study in dosimetry foils, IRDFF-II library (Trkov et al 2020) has been used. The validity of the VR-1 calculation model has been proved in past in criticality experiments (Huml et al., 2013), evaluation of measurement of reaction rates (Rataj et al., 2014,Kostal et al 2021), kinetics parameters (Bily et al., 2019) and neutron spectrum in the radial channel (Kostal et al 2018,Losa et al. 2021).

To improve the convergence of the resulting spectrum and the reaction rates calculated by a track length estimate of the cell flux, a variance reduction method based on the superimposed mesh weight window generation (Werner et al 2017) was employed in the critical calculations.

The calculated VR-1 spectra at the six irradiation points 1(B3), 2(E3), 3(G3), 4(E5), 5(A5) and 6(B8) are shown in Figure 7. The difference in the absolute values of the neutron spectra by two orders of magnitude reflects the variation of the neutron flux inside the VR-1 reactor: the farther from the core center the lower the total flux and absolute energy spectrum. Thus, as seen in Figure 7, the neutron spectrum in channels 6(B8) and 5(A5), which are located outside the core, has lower values. We even could state that the neutron flux in the irradiation channel correlates with number of fission fuel cells which it has nearby. Thus, the flux reaches the minimum value in channel 6(B8) which is separated by one water cell from the fission core and is surrounded by the water.

At first glance the energy shapes of spectrum ratio to $^{235}U(n_{th},f)$ PFNS are rather similar and nearly flat above ≈ 6 MeV for all irradiation locations. However, an increase in the ratio energy dependence by ≈ 5 – 8 % can be found when the secondary neutron energy varies from 3 - 5 MeV to 14 - 16 MeV. The largest energy gradient ≈ 6%/MeV for the ratio of the VR-1 over $^{235}U(n_{th},f)$PFNS shapes is again observed for channels 6(B8) and 5(A5) located outside the fission core.

In our previous work Kostal et al 2021, the origin of such differences in the spectrum was explained by the transmission of the tubular fuel assembly elements, which consists of enriched $UO_2$ fuel in Al claddings surrounded by water. Analytical calculations have proved that the fine structure in the VR-1 spectrum up to ≈ 18 MeV is determined by the fluctuation of the total $^{16}O$ cross section. Whereas the different overall energy trend (gradient) of the VR-1 spectrum and pure $^{235}U(n_{th},f)$ PFNS are likely caused by the change in $^{1}H(n,tot)$ cross-section, which decreases by a factor 2 - 3 in the considered energy range.

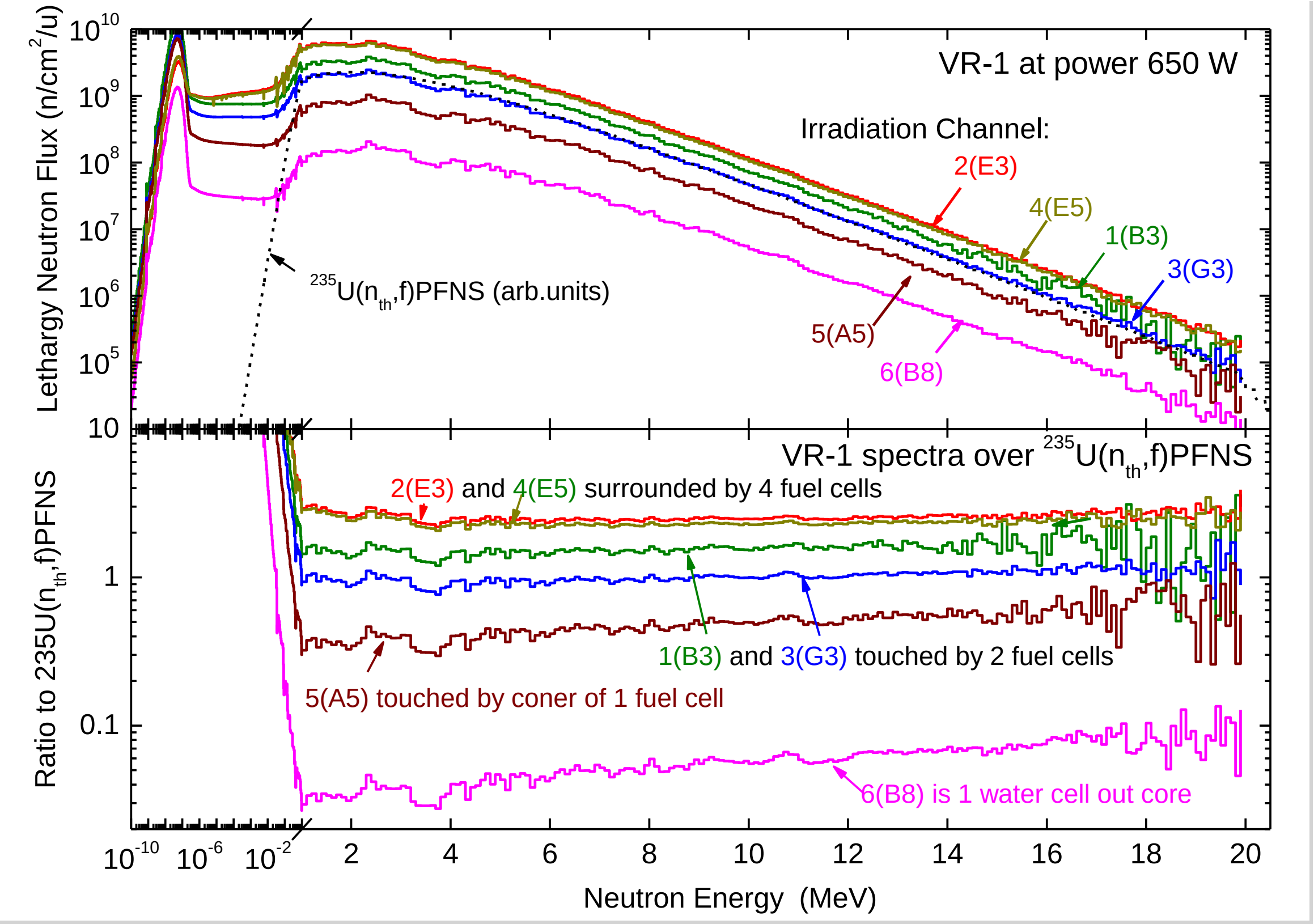

Figure 7: Calculated neutron spectra in the different irradiation channels of the VR-1 reactor and its ratio over $^{235}U(n_{th},f)$ PFNS. The location of the irradiation channels in VR-1 and their labelling are shown in Figure 3 and Table 3.

# 4 Results

## 4.1 Reaction rates ratios

The reaction rate ratio is a dimensionless quantity used for the evaluation of the neutron spectrum shape. It is determined as the ratio of reaction rates of reactions with different thresholds. In this case, it is evaluated as a ratio to $^{27}Al(n,\alpha)$, which was used as a monitoring reaction due to low uncertainties in its cross section. Please also note that the $^{27}Al(n,\alpha)$ reaction is insensitive to additional activation due to the photon-induced reactions as may happen for (n,2n) dosimetry reactions. The calculated differential sensitivity of various reactions is plotted in Figure 8, where the shares of the reaction rate in a separate energy group to the total reaction rate are shown.

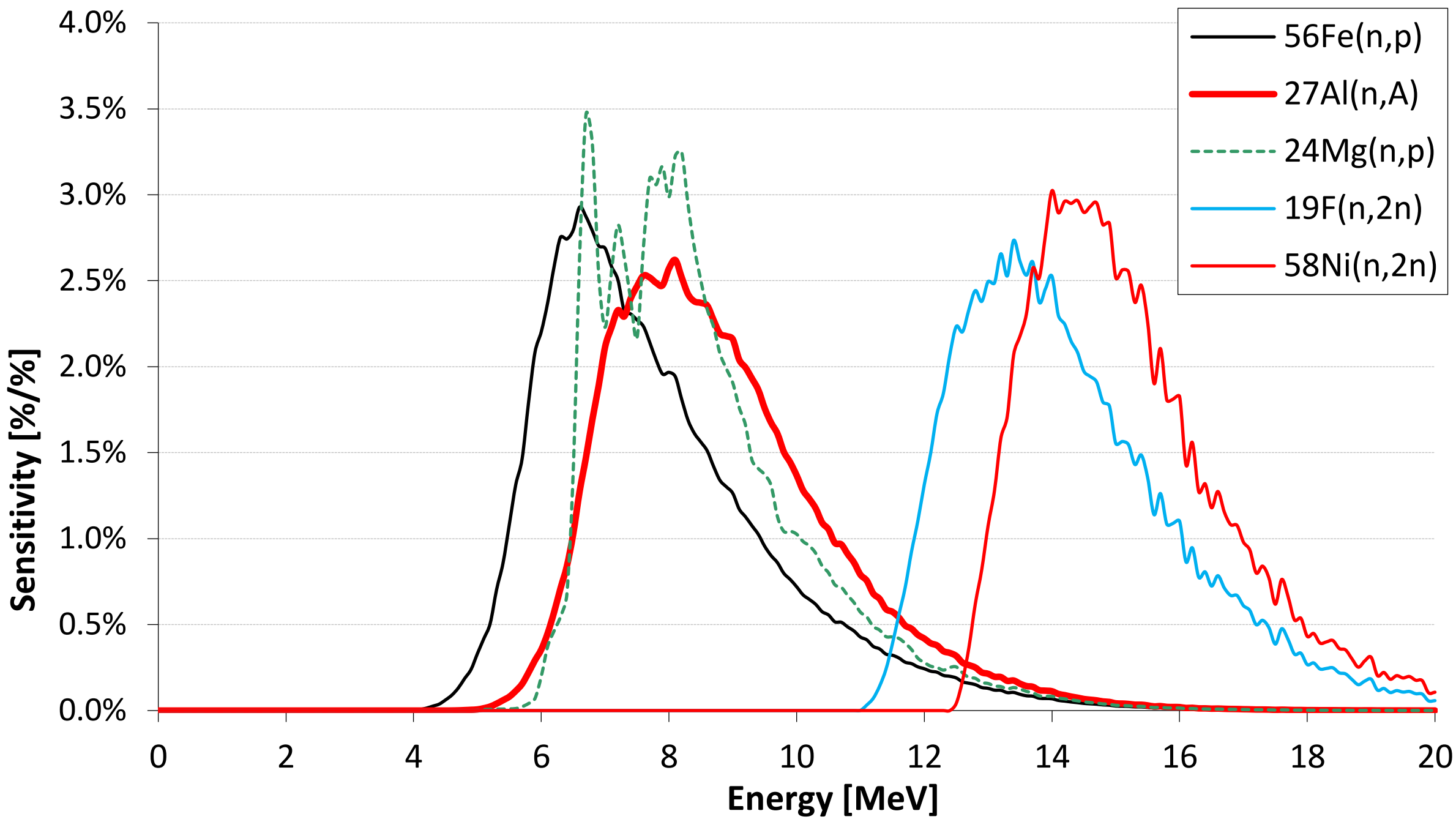

Figure 8: Sensitivity of selected dosimetry reactions

The experimental reaction rates normalized to the value of $^{27}$Al(n,α) SACS averaged over $^{235}$U($n_{th}$,f) PFNS are listed in Table 7. In the neutron fields where the neutron spectrum shape is close to the $^{235}$U($n_{th}$,f) PFNS (Trkov et al 2020), these values are directly SACS averaged over $^{235}$U PFNS. Because similarity between the actual neutron spectra and the $^{235}$U($n_{th}$,f) PFNS can be assumed in the reactor core and positions close to uranium fuel, the normalized averaged cross section was compared to the actual SACS averaged over $^{235}$U($n_{th}$,f) PFNS in last column of Table 7 Trkov et al 2020.

Table 7.: Experimental RR [mb] normalized to $^{27}$Al(n, α) SACS in the $^{235}$U($n_{th}$,f) PFNS (being 0.7005 mb)

| | 1 | 2 | 3 | 4 | 5 | 6 | $^{235}$U(n_th, fiss) PFNS | Unc. |
|---|---|---|---|---|---|---|---|---|
| $^{58}$Ni(n,2n) | 4.640E-3 | 3.990E-3 | 4.526E-3 | 4.057E-3 | 4.963E-3 | 6.091E-3 | 4.076E-3 | 2.98% |
| $^{19}$F(n,2n) | 8.637E-3 | 7.941E-3 | 8.580E-3 | 8.017E-3 | 9.214E-3 | 1.008E-2 | 8.138E-3 | 3.26% |
| $^{197}$Au(n,2n) | 5.087 | 3.460 | 4.499 | 3.432 | 5.784 | 7.236 | 3.387 | 1.67% |
| $^{27}$Al(n,α) | 0.7005 | 0.7005 | 0.7005 | 0.7005 | 0.7005 | 0.7005 | 0.7005 | 1.19% |
| $^{24}$Mg(n,p) | 1.513 | 1.412 | 1.488 | 1.420 | 1.558 | 1.539 | 1.449 | 1.43% |
| $^{48}$Ti(n,p) | 0.2960 | 0.2988 | 0.2938 | 0.2968 | 0.2981 | 0.2892 | 0.3014 | 1.46% |
| $^{56}$Fe(n,p) | 1.083 | 1.097 | 1.088 | 1.094 | 1.074 | 1.041 | 1.079 | 1.54% |
| $^{46}$Ti(n,p) | 10.66 | 10.90 | 10.36 | 10.71 | 9.88 | 9.04 | 11.51 | 1.99% |
| $^{47}$Ti(n,p) | 16.62 | 18.54 | 16.99 | 18.68 | 14.39 | 12.66 | 17.84 | 1.99% |

The comparison is listed in Table 8. It is worth noting that in the central positions (Position 2 and 4), the differences between normalized SACS and SACS averaged over the $^{235}$U($n_{th}$,f) PFNS are negligible, and in the center of non-boundary positioned fuel assemblies, it can be assumed that the spectrum is identical to the $^{235}$U($n_{th}$,f) PFNS.
The discrepancies can be seen in positions at the core boundary (Positions 1 and 3). In higher threshold reactions $^{58}$Ni(n,2n) and $^{19}$F(n,2n), the actual value overpredicts the SACS averaged over $^{235}$U($n_{th}$,f) PFNS. The magnitude of the overprediction is about 11% in the case of

$^{58}Ni(n,2n)$, while in the case of $^{19}F(n,2n)$, it is about 6%. This result reflects the effect of water transmission on neutron spectrum, which hardens the neutron spectrum and increases the relative share of higher energy neutrons (see Kostal et al 2020).
This trend is observed in positions across the water gap as well. At position 6, where the effective water gap is the highest, there is also the highest difference between the measured SACS and the SACS in the $^{235}U(n_{th},f)$ PFNS.

Table 8.: Differences between actual normalized RR and RR in $^{235}U(n_{th},f)$ SACS

| | 1 | 2 | 3 | 4 | 5 | 6 |
|---|---|---|---|---|---|---|
| $^{58}Ni(n,2n)$ | 13.9% | -2.1% | 11.1% | -0.4% | 21.8% | 49.5% |
| $^{19}F(n,2n)$ | 6.2% | -2.4% | 5.5% | -1.5% | 13.3% | 23.9% |
| $^{24}Mg(n,p)$ | 4.5% | -2.5% | 2.7% | -2.0% | 7.6% | 6.3% |
| $^{48}Ti(n,p)$ | -1.8% | -0.8% | -2.5% | -1.5% | -1.1% | -4.0% |
| $^{56}Fe(n,p)$ | 0.4% | 1.7% | 0.8% | 1.4% | -0.4% | -3.5% |
| $^{46}Ti(n,p)$ | -7.4% | -5.3% | -10.0% | -7.0% | -14.1% | -21.4% |
| $^{47}Ti(n,p)$ | -6.8% | 4.0% | -4.7% | 4.8% | -19.3% | -29.0% |

The results of the experimental differences (Figure 9) were compared with the calculational prediction of the same ratio and are listed in Table 9. The trend is comparable to the experimental values only for the magnitudes of the differences, but especially in the case of positions in the core center overestimating the experiment. Based on this result, it can be concluded that in the core center the actual experiment shows that the spectrum is undistinguishable from $^{235}U(n_{th},f)$ PFNS, while the calculation predicts notable hardening of the neutron spectra.

Table 9.: Differences between calculated normalized RR and experimentally obtained RR ratio

| | 1 | 2 | 3 | 4 | 5 | 6 |
|---|---|---|---|---|---|---|
| $^{58}Ni(n,2n)$ | 15.4% | 6.6% | 12.5% | 6.1% | 28.7% | 28.6% |
| $^{19}F(n,2n)$ | 12.0% | 5.3% | 10.3% | 4.7% | 22.4% | 25.1% |
| $^{24}Mg(n,p)$ | -0.6% | -0.2% | -0.5% | -0.3% | -1.0% | -1.4% |
| $^{48}Ti(n,p)$ | -0.5% | -0.1% | -0.4% | -0.1% | -0.9% | -1.4% |
| $^{56}Fe(n,p)$ | -1.9% | -0.6% | -1.7% | -0.5% | -3.4% | -4.9% |
| $^{46}Ti(n,p)$ | -5.2% | -0.9% | -4.5% | -0.5% | -10.5% | -15.0% |
| $^{47}Ti(n,p)$ | -6.2% | 3.7% | -4.5% | 5.1% | -18.1% | -26.8% |

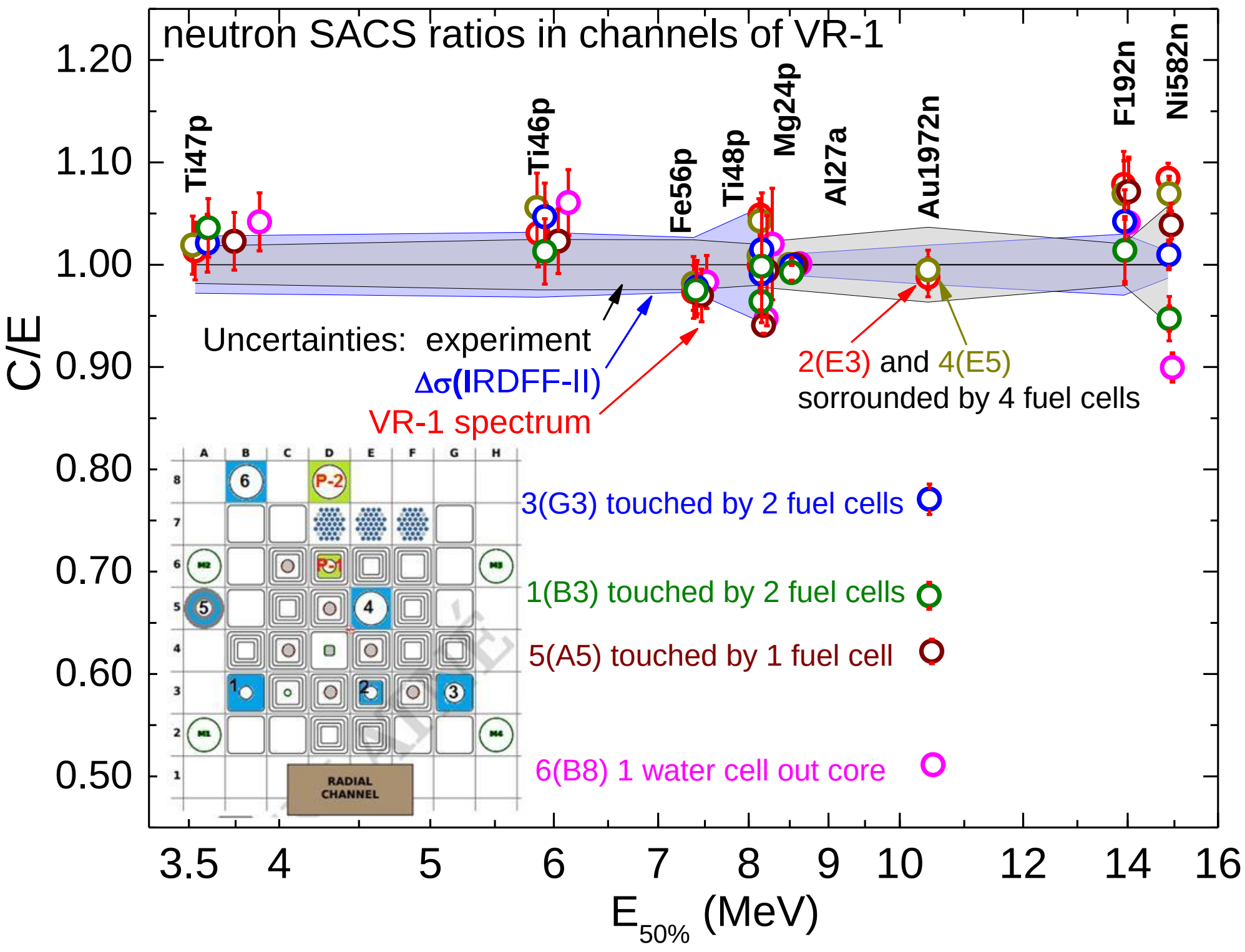

Figure 9: The C/E ratio for the SACS of neutron dosimetry reactions calculated with IRDFF-II library. The insert shows the location of six foils in the VR-1 channels and their labelling. The measurement uncertainties are shown by grey corridor, the contribution from IRDFF-II cross sections – blue bars, from calculated VR-1 neutron spectrum – red bars.

### 4.2 Spectrum averaged neutron cross sections and problem of gold in the VR-1 outcore channels

The spectrum average neutron dosimetry cross sections (SACS) in the MCNP simulated neutron spectra shown in Figure 7 were calculated by code RR_UNC (Trkov et al 2001). The neutron dosimetry cross sections and their covariances were taken from the IRDFF-II library (Trkov et al 2020). The experimental and calculated results for the eight dosimetry reactions at six irradiation positions in VR-1 are summarized in Table 10 and graphically intercompared in Figure 9. The absolute values and uncertainties for experimental SACS were derived from reaction rates listed in Table 5 and 6 after normalization to the monitoring reaction $^{27}Al(n,\alpha)^{24}Na$ independently in every irradiation channel. Consequently, the C/E ratio for this reference reaction equals to unity. It is worth noting that the SACS uncertainties resulting from the MCNP simulation of the VR-1 spectra are smaller than those propagated from the IRDFF-II cross sections, even for the high threshold reactions.

For most reactions, we observe an agreement within the estimated experimental and calculated uncertainties. Exception is the reaction $^{197}Au(n,2n)^{196}Au$, but only for irradiation in four positions: 1(B3), 3(G3), 5(A5) and 6(B8), where the reaction rate underestimation reaches 20 – 50%. A moderate 10% underestimation is also observed for $^{58}Ni(n,2n)$ but only in channel 6(B8).

Table 10.: The SACS values and their uncertainties measured relative to $^{27}Al(n,\alpha)$ in the six irradiation channels of the VR-1 reactor. C/E ratios and total uncertainties $\Delta(C/E)$ computed with cross sections from IRDFF-II

| Reaction | $E_{50\%}$ [MeV] | 1 (B3) | | | | 2 (E3) | | | | 3 (G3) | | | |
|---|---|---|---|---|---|---|---|---|---|---|---|---|---|
| | | SACS [mb] | Unc. % | C/E | Δ(C/E) [%] | SACS [mb] | Unc. [%] | C/E | Δ(C/E) [%] | SACS [mb] | Unc. [%] | C/E | Δ(C/E) [%] |
| $^{47}Ti(n,p)^{47}Sc$ | 3.60 | 2.59E+00 | 1.8 | 1.044 | 3.3 | 5.44E+00 | 2.4 | 1.013 | 3.3 | 2.70E+00 | 1.8 | 1.022 | 3.3 |
| $^{46}Ti(n,p)^{46}Sc$ | 5.92 | 1.66E+00 | 2.6 | 1.021 | 4.1 | 3.20E+00 | 2.5 | 1.031 | 4.0 | 1.65E+00 | 2.2 | 1.048 | 3.8 |
| $^{56}Fe(n,p)^{56}Mn$ | 7.41 | 1.68E-01 | 3.2 | 0.983 | 4.2 | 3.23E-01 | 2.0 | 0.973 | 3.7 | 1.73E-01 | 2.6 | 0.979 | 3.7 |
| $^{48}Ti(n,p)^{48}Sc$ | 8.16 | 4.61E-02 | 2.2 | 1.006 | 5.9 | 8.77E-02 | 2.1 | 0.999 | 5.9 | 4.68E-02 | 1.8 | 1.016 | 5.8 |
| $^{24}Mg(n,p)^{24}Na$ | 8.15 | 2.36E-01 | 2.1 | 0.972 | 2.3 | 4.13E-01 | 3.7 | 1.048 | 2.3 | 2.37E-01 | 2.1 | 0.992 | 2.3 |
| $^{197}Au(n,2n)^{196}Au$ | 10.45 | 7.93E-01 | 6.8 | 0.682 | 7.1 | 1.02E+00 | 1.9 | 0.988 | 4.2 | 7.18E-01 | 5.4 | 0.772 | 5.7 |
| $^{19}F(n,2n)^{18}F$ | 13.94 | 1.34E-03 | 1.9 | 1.022 | 3.8 | 2.33E-03 | 5.3 | 1.078 | 3.6 | 1.37E-03 | 2.0 | 1.043 | 3.6 |
| $^{58}Ni(n,2n)^{57}Ni$ | 14.88 | 7.29E-04 | 3.3 | 0.955 | 4.0 | 1.17E-03 | 2.0 | 1.085 | 5.5 | 7.18E-04 | 5.3 | 1.011 | 5.5 |

| Reaction | $E_{50\%}$ [MeV] | 4 (E5) | | | | 5 (A5) | | | | 6 (B8) | | | |
|---|---|---|---|---|---|---|---|---|---|---|---|---|---|
| | | SACS [mb] | Unc. % | C/E | Δ(C/E) [%] | SACS [mb] | Unc. [%] | C/E | Δ(C/E) [%] | SACS [mb] | Unc. [%] | C/E | Δ(C/E) [%] |
| $^{47}Ti(n,p)^{47}Sc$ | 3.60 | 5.15E+00 | 1.8 | 1.019 | 3.3 | 1.76E+00 | 1.8 | 1.023 | 3.3 | 1.91E+00 | 1.8 | 1.041 | 3.3 |
| $^{46}Ti(n,p)^{46}Sc$ | 5.92 | 2.94E+00 | 2.2 | 1.056 | 3.9 | 1.21E+00 | 3.0 | 1.023 | 4.3 | 1.36E+00 | 3.0 | 1.060 | 4.3 |
| $^{56}Fe(n,p)^{56}Mn$ | 7.41 | 3.01E-01 | 2.6 | 0.982 | 3.7 | 1.31E-01 | 4.2 | 0.970 | 5.0 | 1.57E-01 | 3.3 | 0.982 | 4.2 |
| $^{48}Ti(n,p)^{48}Sc$ | 8.16 | 8.15E-02 | 2.2 | 1.009 | 6.0 | 3.65E-02 | 2.5 | 0.994 | 6.0 | 4.36E-02 | 1.8 | 1.019 | 5.6 |
| $^{24}Mg(n,p)^{24}Na$ | 8.15 | 3.90E-01 | 2.1 | 1.043 | 2.3 | 1.91E-01 | 2.3 | 0.941 | 2.4 | 2.32E-01 | 2.6 | 0.947 | 2.7 |
| $^{197}Au(n,2n)^{196}Au$ | 10.45 | 9.44E-01 | 4.1 | 0.995 | 4.5 | 7.07E-01 | 5.9 | 0.622 | 6.2 | 1.09E+00 | 6.8 | 0.511 | 7.1 |
| $^{19}F(n,2n)^{18}F$ | 13.94 | 2.21E-03 | 2.0 | 1.069 | 3.7 | 1.13E-03 | 2.0 | 1.071 | 3.7 | 1.52E-03 | 2.1 | 1.040 | 3.7 |
| $^{58}Ni(n,2n)^{57}Ni$ | 14.88 | 1.12E-03 | 2.8 | 1.069 | 3.2 | 6.07E-04 | 3.3 | 1.039 | 3.9 | 9.18E-04 | 2.0 | 0.899 | 2.6 |

The reason of up to 50% underestimation of the neutron SACS for reaction $^{197}Au(n,2n)^{196}Au$ could be an impact of the photo-nuclear reaction $^{197}Au(\gamma,n)^{196}Au$ which leads to the additional production of the same residual $^{196}Au$. To demonstrate the principal possibility of such effect we plotted in Figure 10 the prompt γ-ray energy spectra from the thermal neutron capture on $^{nat}Fe$ (main structural element of the VR1 core) and from the thermal-neutron induced fission of $^{235}U$ (the fissile fuel). The evaluated data were taken from ENDF/B-VIII.0 (Brown et al, 2018, Stetcu et al., 2020). The same figure also displays the cross sections for several (γ,n) reactions, which may compete with (n,2n) neutron dosimetry reactions. The displayed (γ,n) data are taken from the recent IAEA photo-nuclear data library IAEA/PD-2019 (Kawano et al. 2020). It is seen that the γ-ray spectrum from $^{nat}Fe(n,\gamma)$ is substantially harder than γ-ray spectrum from $^{235}U(n,f)$ (being $^{238}U$ generally irrelevant for thermal reactors) and overlaps only with cross section $^{197}Au(\gamma,n)^{196}Au$ above its threshold 8.073 MeV. **Chyba! Nenalezen zdroj odkazů.** shows that thermal capture on two minor iron isotopes $^{54}Fe$ and $^{57}Fe$ can produce γ-rays with energies above 8.073 MeV and thus could be a dominant source of high-energy neutrons via the corresponding (γ,n) reactions.

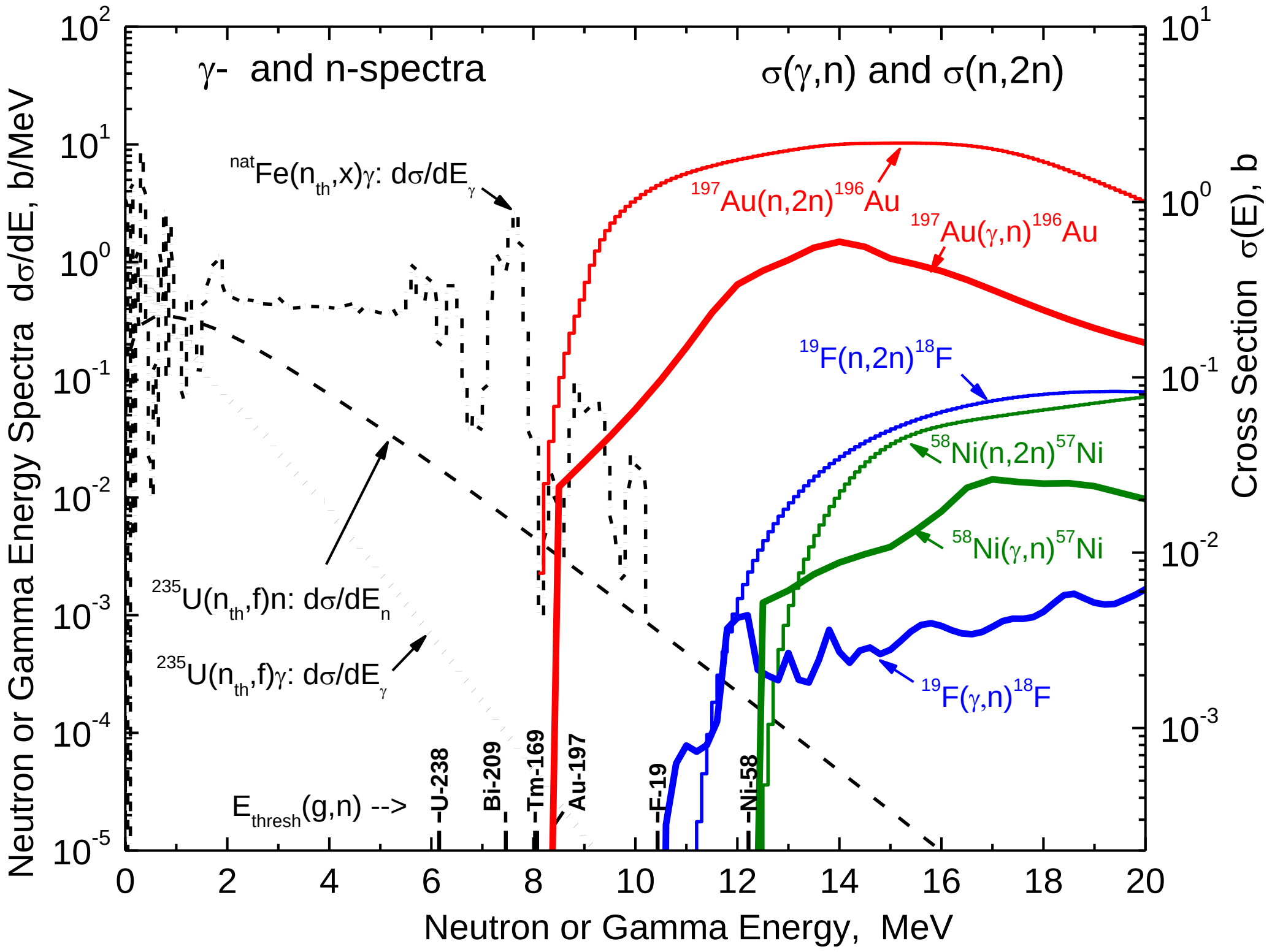

Figure 10: Left axis: spectra of γ-rays and prompt neutrons from thermal neutron induced reactions on $^{nat}Fe$ and $^{235}U$ taken from ENDF/B-VIII.0. Right axis: cross sections for neutron dosimetry reactions from IRDFF-II (histograms) and photo-nuclear reactions from IAEA/PD-2019 (thick curves). The kinematic thresholds for several (γ,n) reactions competing with corresponding IRDFF (n,2n) reactions are identified just above the X-axis.

From this qualitative consideration we conclude that the prompt γ-rays produced by capture of the thermal neutrons in the minor iron isotopes will probably generate $^{196}Au$ via the $^{197}Au(\gamma,n)$ reaction additionally to the conventional neutron path $^{197}Au(n,2n)$. The fission reactions $^{235}U(n,f)\gamma$ and $^{238}U(n,f)\gamma$ may also contribute since their γ–ray energies extend up to ≈ 21 MeV, however those prompt fission γ-yields exponentially decrease with outgoing photon energy. Inelastic scattering reactions probably will be a minor source of high energy gammas due to the much lower inelastic scattering cross sections compared to thermal capture and fission cross sections. This is especially true of inelastic scattering on heavy elements including actinides, where inelastic scattering gammas can be neglected compared to prompt fission gammas (Stetcu et al., 2020).

Table 11.: Maximum energies of γ–rays from neutron induced capture (n,γ), fission (n,f) and inelastic scattering (n,n') for the main structural and fissile isotopes presented in the VR-1 reactor. The γ–ray energy threshold for the several photo-nuclear reactions which compete with the dosimetry neutron reactions utilized in the present work.

| Maximum Eγ from the neutron induced reactions | | | | $E_{thr}$ of photo-nuclear reactions | |
|---|---|---|---|---|---|
| Isotope | (n,γ) MeV | (n,f) MeV | (n.n') MeV | Target Isotope | (γ,n) MeV |
| $^{16}O$ | 4.143 | | ≤ 20 | $^{197}Au$ | 8.072 |
| $^{27}Al$ | 7.725 | | ≤ 20 | $^{19}F$ | 10.432 |
| $^{54}Fe$ | 9.298 | | ≤ 20 | $^{23}Na$ | 12.216 |
| $^{56}Fe$ | 7.646 | | ≤ 20 | | |
| $^{57}Fe$ | 10.044 | | ≤ 20 | | |
| $^{58}Fe$ | 6.580 | | ≤ 20 | | |
| $^{235}U$ | 6.544 | 21.250 | ≤ 20 | | |
| $^{238}U$ | 4.830 | 21.250 | ≤ 20 | | |

### 4.3 Calculated γ-ray spectra in the VR-1 irradiation channels and γ-ray induced SACS

The γ-ray energy spectra at the dosimetry foil irradiation positions were accurately simulated by the MCNP code employing the ENDF/B-VIII.0 neutron-photon transport data, i.e. the same data as for the neutron spectrum calculations. As an example, the γ-ray spectra in the irradiation channels 6(B8) (separated by one water cell from the fission core) and 2(E3) (inside the core, i.e., surrounded by fuel elements from all sides) are shown in Figure 11. Indeed, one observes the discrete γ-lines below 10 MeV, which obviously stem from reaction Fe(n,γ), and above - the smooth γ-spectrum extending up to 20 MeV, which is produced by reaction U(n,f)γ and high energy neutron inelastic scattering (the fluctuation above ≈ 15 MeV are due to insufficient Monte Carlo simulation statistics).

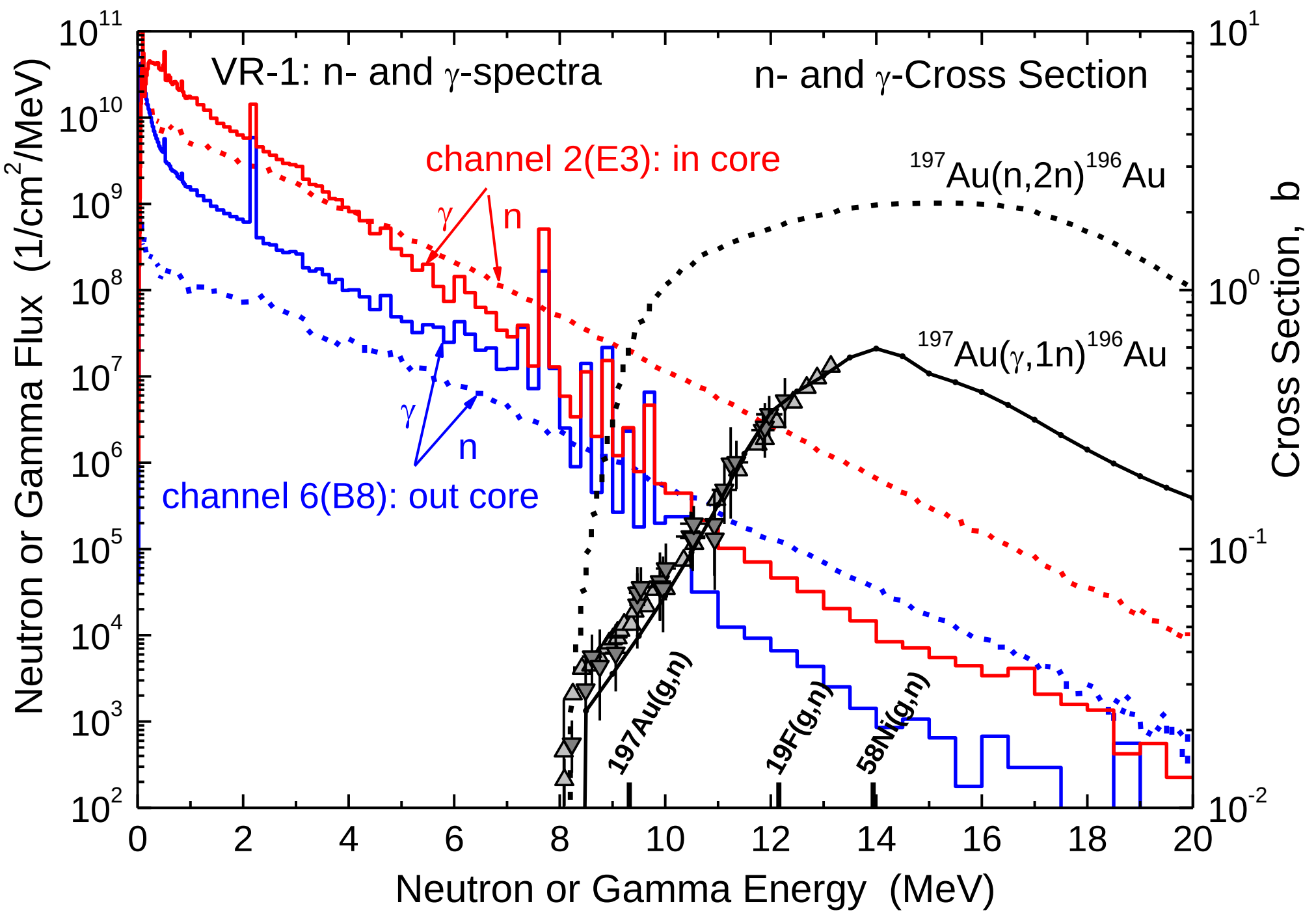

Figure 11: Left axis: spectra of γ-rays and prompt neutrons in the irradiation channels 3(E3) located in the VR-1 fission core and 6(B8) located out of core in the water. Right axis: cross sections for neutron dosimetry reactions from IRDFF-II (dashed curve) and for photo-nuclear reactions from IAEA/PD-2019 (solid curve). Experimental cross section for $^{197}$Au(γ,n) are from Hara et al. 2007 and Itoh et al. 2011 (grey symbols). The γ-ray energies, where the γ-SACS integrals reach 50%, are indicated by vertical bars for three reactions $^{197}$Au(γ,n), $^{19}$F(γ,n) and $^{58}$Ni(γ,n).

The corresponding γ-ray induced spectrum averaged cross sections, γ-SACS, were computed by the RR_UNC code utilizing the (γ,n) cross sections from the photo-nuclear data library IAEA/PD-2019 (Kawano et al. 2020). The normalizations of the γ-ray and neutron fluxes as well as corresponding reaction rates was performed per one fission neutron in the VR-1 fissile core (i.e., per one neutron in the MCNP kcode regime). The results are listed and inter-compared in Table 12. The (γ,n) contribution fraction is defined as a ratio γ-SACS / (n-SACS + γ-SACS). It is seen that the neutron dosimetry reaction (n,2n) experiences the maximum competition from the photo-nuclear reaction (γ,n) in the case of the gold dosimeter. The effect is most substantial in the irradiation channel 6(B8): $^{197}$Au(γ,n) contributes 11%, whereas $^{197}$Au(n,2n) - 89% to the total production of $^{196}$Au. The channel 6(B8) is separated from the VR-1 fissile core by one cell filled with water. The latter, as seen in Table 13 and Figure 11, attenuates the neutron flux by approximately one order of magnitude, however the γ-ray flux drops only by factor 3. This increases the balance in favor of γ-rays and hence the number of the (γ,n) events in comparison with (n,2n). In other five irradiation channels the (γ,n)/(n,2n) fraction for $^{197}$Au varies between much smaller values 0.4 and 1.2%. The γ/n competition for other studied isotopes $^{19}$F and $^{58}$Ni is also small since they have higher (γ,n) thresholds or 50% response energy $E_{50\%}$ as seen in Table 11 and Table 12. The fraction (γ,n)/(n,2n) is larger for $^{58}$Ni than for $^{19}$F and ranges from 0.4 to 2.0% reaching the maximum value once again in the outcore channel 6(B8).
The γ-ray energies $E_{50\%}$, where the γ-SACS integral reaches 50%, are listed in Table 12 and plotted in Figure 11 for three reactions $^{197}$Au(γ,n), $^{19}$F(γ,n) and $^{58}$Ni(γ,n). It is interesting to observe that the $^{197}Au(\gamma,n)$ SACS is sensitive to the γ-ray spectra in the energy interval 8.5 – 10.5 MeV where the discrete (primary) γ-rays are clearly manifest themselves (see explanation of their origin in the next section). Another two reactions $^{19}F(\gamma,n)$ and $^{58}Ni(\gamma,n)$ are sensitive to the smooth part of γ-spectrum from 11 to 15 MeV, where the (n,f)γ and (n,n')γ reactions contribute. Due to accumulated Monte Carlo statistics for the γ-ray spectra in this energy domain the uncertainties of γ-SACS for $^{19}F$ and $^{58}Ni$ do not exceed (6 –15)% in the worst case.

Applying corrections for the contribution of the photo-nuclear reaction $^{197}Au(\gamma,n)^{196}Au$ to the ratio C/E{n} = 0.512, obtained in the outcore irradiation channel 6(B8) only with neutron induced reaction $^{197}Au(n,2n)^{196}Au$, results to the ratio in the mixed neutron-gamma field C/E{n+γ} = 0.573. It has improved an agreement between calculations and measurements, but a 42% underestimation still exists.

### 4.1 Validation of the $^{nat}Fe(n_{th},x)\gamma$ spectra presented in various versions of the ENDF/B library

To investigate the potential reasons why our calculations still underestimate the experimentally observed production of $^{196}Au$ in the several VR-1 irradiation channels, we have analysed the status of the measured and evaluated $^{nat}Fe(n_{th},x)\gamma$ ray spectra. The experimental γ-ray spectra were taken from the adopted files of the PGAA library released in year 2007 (PGAA 2007 ). It has to be noticed that we replaced the PGAA data for $^{56}Fe(n,\gamma)^{57}Fe$ by the recent measurement and evaluation of R. Firestone et al. published in 2017 (Firestone et al. 2017) (that however had no notable impact on our calculational results). Furthermore, for reaction $^{57}Fe(n,\gamma)^{58}Fe$, two γ-transitions 9.233 and 10.044 MeV not included in the adopted PGAA files, were added from 2001 evaluation of R. Reedy and S. Frankle (Franke et al. 2001). The γ-ray spectra for $^{nat}Fe(n_{th},x)\gamma$ reaction were extracted from three latest versions of the evaluated data library ENDF/B with the help of the MCNP code.

Table 12.: Competition between (n,2n) and (γ,n) reactions leading to the same residual nucleus in the six VR-1 irradiation channels (here they are ordered from location inside the core to the periphery). For every channel the energy integrated neutron and γ-ray fluxes and their ratio are additionally listed. Columns contain the neutron and γ-ray 50% response energies, SACS with total relative uncertainties, reaction rates and production fraction of the reaction residuals.

| Reaction | $E_{50\%}$, MeV | SACS, mb | Rel. Unc., % | Reaction Rate, 1/kcode | Fraction, % |
|---|---|---|---|---|---|
| Channel 2(E3) is surrounded by 4 fuel cells: n-flux = 6.405E-4 n/cm$^2$, γ-flux = 8.187E-4 γ/cm$^2$, γ/n = 1.278 | | | | | |
| $^{197}$Au(n,2n)$^{196}$Au | 10.427 | 1.002E+00 | 1.93 | 6.418E-07 | 99.39 ± 0.01 |
| $^{197}$Au(γ,n)$^{196}$Au | 9.573 | 4.841E-03 | 0.60 | 3.963E-09 | 0.61 ± 0.01 |
| $^{19}$F(n,2n)$^{18}$F | 13.919 | 2.513E-03 | 3.00 | 1.609E-09 | 99.57 ± 0.02 |
| $^{19}$F(γ,n)$^{18}$F | 12.070 | 8.494E-06 | 2.70 | 6.953E-12 | 0.43 ± 0.02 |
| $^{58}$Ni(n,2n)$^{57}$Ni | 14.862 | 1.270E-03 | 1.36 | 8.137E-10 | 99.02 ± 0.06 |
| $^{58}$Ni(γ,n)$^{57}$Ni | 14.848 | 9.879E-06 | 6.18 | 8.087E-12 | 0.98 ± 0.06 |
| Channel 4(E5) is surrounded by 4 fuel cells: n-flux = 6.302E-4 n/cm$^2$, γ-flux = 8.477E-4 γ/cm$^2$, γ/n = 1.345 | | | | | |
| $^{197}$Au(n,2n)$^{196}$Au | 10.427 | 9.395E-01 | 1.93 | 5.921E-07 | 98.76 ± 0.02 |
| $^{197}$Au(γ,n)$^{196}$Au | 9.271 | 8.735E-03 | 0.36 | 7.405E-09 | 1.24 ± 0.02 |
| $^{19}$F(n,2n)$^{18}$F | 13.927 | 2.358E-03 | 3.06 | 1.486E-09 | 99.51 ± 0.02 |
| $^{19}$F(γ,n)$^{18}$F | 12.048 | 8.716E-06 | 2.63 | 7.389E-12 | 0.49 ± 0.02 |
| $^{58}$Ni(n,2n)$^{57}$Ni | 14.883 | 1.193E-03 | 1.62 | 7.519E-10 | 98.88 ± 0.07 |
| $^{58}$Ni(γ,n)$^{57}$Ni | 14.783 | 1.005E-05 | 5.98 | 8.522E-12 | 1.12 ± 0.07 |
| Channel 1(B3) touches 2 fuel cells: n-flux = 7.551E-4 n/cm$^2$, γ-flux = 5.007E-4 γ/cm$^2$, γ/n = 0.663 | | | | | |
| $^{197}$Au(n,2n)$^{196}$Au | 10.446 | 5.408E-01 | 1.95 | 4.084E-07 | 99.63 ± 0.01 |
| $^{197}$Au(γ,n)$^{196}$Au | 9.617 | 3.029E-03 | 0.97 | 1.517E-09 | 0.37 ± 0.01 |
| $^{19}$F(n,2n)$^{18}$F | 13.943 | 1.373E-03 | 3.26 | 1.037E-09 | 99.70 ± 0.02 |
| $^{19}$F(γ,n)$^{18}$F | 12.022 | 6.249E-06 | 3.91 | 3.129E-12 | 0.61 ± 0.02 |
| $^{58}$Ni(n,2n)$^{57}$Ni | 14.890 | 6.956E-04 | 2.30 | 5.252E-10 | 99.33 ± 0.06 |
| $^{58}$Ni(γ,n)$^{57}$Ni | 15.200 | 7.066E-06 | 9.37 | 3.538E-12 | 0.67 ± 0.06 |
| Channel 3(G3) touches 2 fuel cells: n-flux = 4.71,E-4 n/cm$^2$, γ-flux = 4.766E-4 γ/cm$^2$, γ/n = 1.012 | | | | | |
| $^{197}$Au(n,2n)$^{196}$Au | 10.447 | 5.543E-01 | 1.92 | 2.610E-07 | 99.19 ± 0.02 |
| $^{197}$Au(γ,n)$^{196}$Au | 9.392 | 4.456E-03 | 0.75 | 2.124E-09 | 0.81 ± 0.02 |
| $^{19}$F(n,2n)$^{18}$F | 13.940 | 1.427E-03 | 3.00 | 6.719E-10 | 99.54 ± 0.02 |
| $^{19}$F(γ,n)$^{18}$F | 12.007 | 6.562E-06 | 3.83 | 3.127E-12 | 0.46 ± 0.02 |
| $^{58}$Ni(n,2n)$^{57}$Ni | 14.879 | 7.261E-04 | 1.31 | 3.420E-10 | 98.94 ± 0.10 |
| $^{58}$Ni(γ,n)$^{57}$Ni | 14.718 | 7.663E-06 | 8.91 | 3.652E-12 | 1.06 ± 0.10 |
| Channel 5(A5) touches corner of 1 fuel cells: n-flux = 3.017E-4 n/cm$^2$, γ-flux = 3.382E-4 γ/cm$^2$, γ/n = 1.121 | | | | | |
| $^{197}$Au(n,2n)$^{196}$Au | 10.473 | 4.319E-01 | 1.93 | 1.303E-07 | 98.76 ± 0.03 |
| $^{197}$Au(γ,n)$^{196}$Au | 9.301 | 4.851E-03 | 0.72 | 1.641E-09 | 1.24 ± 0.03 |
| $^{19}$F(n,2n)$^{18}$F | 13.973 | 1.165E-03 | 3.14 | 3.513E-10 | 99.51 ± 0.03 |
| $^{19}$F(γ,n)$^{18}$F | 12.071 | 5.095E-06 | 5.05 | 1.723E-12 | 0.49 ± 0.03 |
| $^{58}$Ni(n,2n)$^{57}$Ni | 14.922 | 5.979E-04 | 2.02 | 1.804E-10 | 98.86 ± 0.14 |
| $^{58}$Ni(γ,n)$^{57}$Ni | 14.674 | 6.172E-06 | 12.47 | 2.087E-12 | 1.14 ± 0.14 |
| Channel 6(B8) behind 1 water cell: n-flux = 5.754E-5 n/cm$^2$, γ-flux = 1.735E-4 γ/cm$^2$, γ/n = 3.016 | | | | | |
| $^{197}$Au(n,2n)$^{196}$Au | 10.507 | 5.576E-01 | 1.90 | 3.208E-08 | 89.39 ± 0.20 |
| $^{197}$Au(γ,n)$^{196}$Au | 8.990 | 2.194E-02 | 0.32 | 3.807E-09 | 10.61 ± 0.20 |
| $^{19}$F(n,2n)$^{18}$F | 14.019 | 1.582E-03 | 2.99 | 9.103E-11 | 98.94 ± 0.08 |
| $^{19}$F(γ,n)$^{18}$F | 12.018 | 5.624E-06 | 6.44 | 9.759E-13 | 1.06 ± 0.08 |
| $^{58}$Ni(n,2n)$^{57}$Ni | 14.960 | 8.249E-04 | 1.59 | 4.746E-11 | 98.01 ± 0.31 |
| $^{58}$Ni(γ,n)$^{57}$Ni | 14.738 | 5.565E-06 | 15.34 | 9.657E-13 | 1.99 ± 0.31 |

The derived evaluated and experimental data are compared in Figure 12. For the more convenient visualization the γ-ray spectra were folded with Gaussian distribution, that imitates the energy resolution ≈ 3%. This Figure also displays the photo-nuclear (γ,n) cross sections for nuclei $^{197}$Au, $^{19}$F and $^{58}$Ni from IAEA/PD-2019. It is seen that only $^{197}$Au(γ,n)$^{196}$Au reaction which has the minimal threshold 8.073 MeV will occur in the $^{nat}$Fe($n_{th}$,x) γ-ray spectrum.

Moreover it will be triggered only by the highest energy γ-rays which are primary transitions to the ground states in following reactions $^{54}Fe(n_{th},\gamma_0)^{55}Fe$(g.s.) and $^{57}Fe(n_{th},\gamma_0)^{58}Fe$(g.s.).

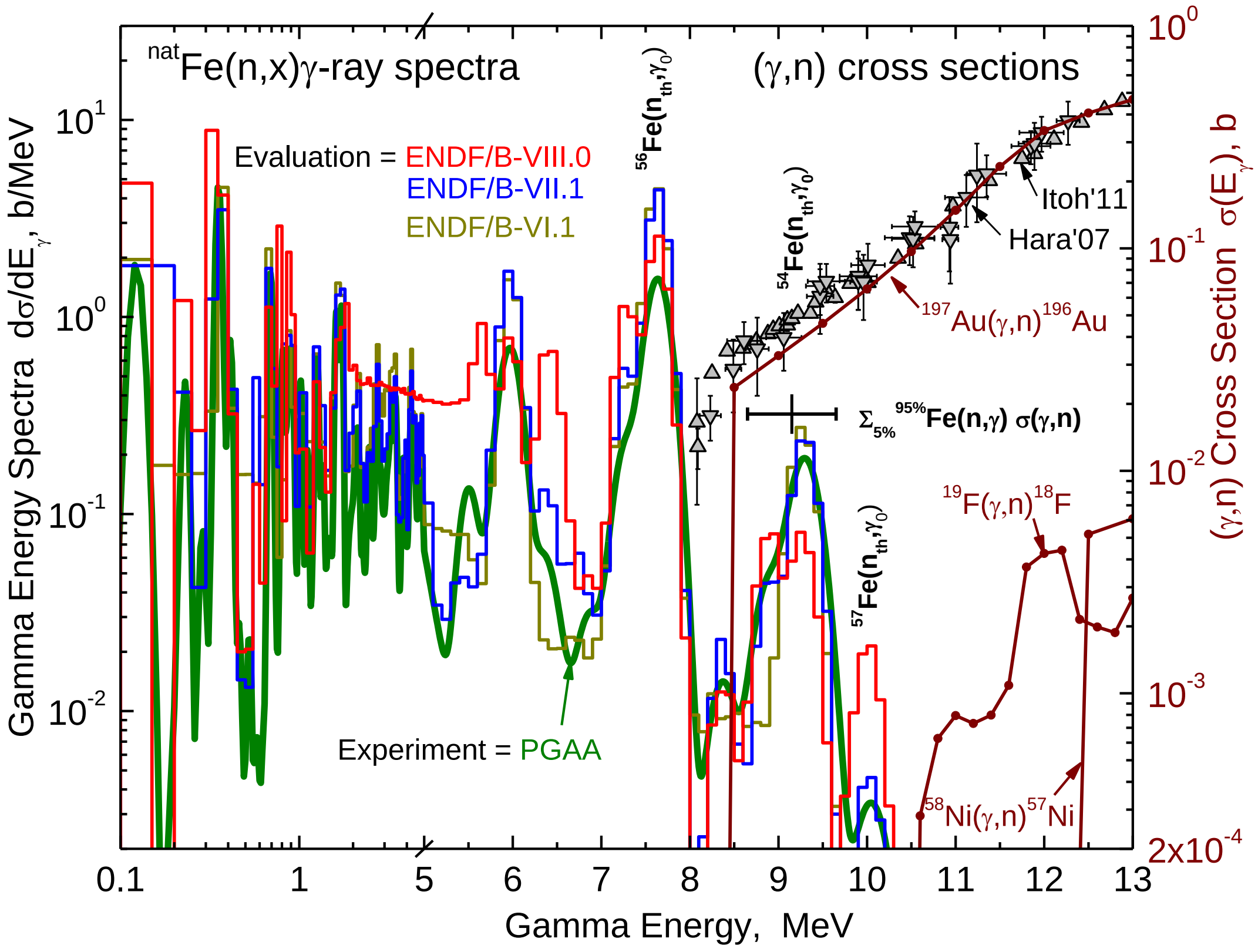

Figure 12: (Left axis) Spectra of prompt γ-rays from the thermal neutron induced capture on $^{nat}Fe$ obtained from ENDF/B-VIII.0, ENDF/B-VII.1 and ENDF/B-VI.1 in comparison with experimental γ-spectra taken from PGAA (PGAA 2007 ) and Firestone (Firestone et al. 2017). The position of the $\gamma_0$ from $^{54}Fe(n_{th},\gamma_0)^{55}Fe$(g.s.) and $^{57}Fe(n_{th},\gamma_0)^{5}Fe$(g.s.) are identified by corresponding labels. (Right axis) Cross section for the photo-nuclear reaction (γ,n) on $^{197}Au$, $^{58}Ni$ and $^{19}F$ taken from IAEA/PD-2019. Experimental cross sections for $^{197}Au$(γ,n) are from Hara et al. 2007 and Itoh et al. 2011. The (5 – 95)% response energy interval of the product $^{nat}Fe$(n,x)γ and σ(γ,n) is shown by the horizontal bars. *Note the change of the X-axis scale at 5 MeV from log to linear.*

Table 13 lists the γ-SACS obtained by integration of the $^{nat}Fe(n_{th},x)\gamma$ ray spectra and $^{196}Au$(γ,n) cross section product, after normalization to the Maxwellian thermal capture averaged cross section of 2.567 b (Mughabghab 2018). The γ-ray energy at which the integral reaches 50%, $E_{\gamma 50\%}$ = 9.2 MeV, as well as the (5 – 95)% energy response interval $E_{\gamma 5\%}$ - $E_{\gamma 95\%}$ = (9.0-9.5) MeV definitely show that the $^{197}Au$(γ,n) SACS is sensitive only to $^{54}Fe(n_{th},\gamma_0)^{55}Fe$(g.s.). The C/E ratio, defined as γ-SACS(ENDF)/γ-SACS(PGAA), indicates that in the overlapping energy range from 8.1 to 10.3 MeV the earlier versions ENDF/B-VII.1 and ENDF/B-VI.1 do better reproduce the $^{nat}Fe(n_{th},x)\gamma$-ray spectrum than the latest one. Figure 12 also shows that the thermal capture γ-rays from iron and other VR-1 structural materials which are listed in Table 11 will have no impact on the $^{19}F$ and $^{58}Ni$ γ-SACS.
The ENDF/B-VIII.0 underestimates the yield of γ-rays by 40 %. Since this version was used in the present simulation of the γ-ray fields in the VR-1 reactor, it could be a reason of the underestimation of the observed contribution of the photo-nuclear process $^{197}Au$(γ,n) to the production of $^{196}Au$ additionally to $^{197}Au(n,2n)^{196}Au$. Applying the correction C/E{PGAA} =

0.587 found from comparison of ENDF/B-VIII.0 with PGAA we will get C/E{n+γ} = 0.615. This further improves an agreement between our calculations and measurements presented in the paper. The left underestimation 38% we attribute to the $^{197}$Au(γ,n) cross section near the threshold. As seen in Figure 12 the evaluation IAEA/PD-2019 in the energy interval (8.1 – 10.0) MeV has only 2-3 points and was derived mostly from model calculations guided by experimental data (Kawano et al. 2020). To check the reliability of the $^{197}$Au(γ,n) cross section just above its threshold we compared it with known experimental data Hara et al. 2007 and Itoh et al. 2011. As it is clearly seen in Figure 12 the IAEA/PD-2019 evaluation indeed underestimates these experimental data: at key energy Eγ ≈ 9.3 MeV - by about 30%. A better evaluation of the $^{197}$Au(γ,n) cross section based on a least-square fit of measured data could significantly improve our C/E.

Table 13.: The calculated 50%, 5% and 95% response energies for integration of the $^{197}$Au(n,γ) cross section and γ-SACS in the $^{nat}$Fe($n_{th}$,x)γ-spectrum taken from PGAA (experiment) and from three versions of the ENDF/B library. Ratio C/E is γ-SACS(ENDF/B) / γ-SACS (PGAA).

| Source of the $^{nat}$Fe($n_{th}$,x)γ-ray spectrum | $E_{\gamma 50\%}$ MeV | $E_{\gamma 5\%}$ - $E_{\gamma 95\%}$ MeV | γ-SACS mb | C/E |
|---|---|---|---|---|
| PGAA (experiment) | 9.28 | 9.03 – 9.45 | 1.473 | |
| ENDF/B-VIII.0 | 9.15 | 8.75 – 9.70 | 0.864 | 0.587 |
| ENDF/B-VII.1 | 9.15 | 9.00 – 9.30 | 1.409 | 0.957 |
| ENDF/B-VI.1 | 9.15 | 9.00 – 9.20 | 1.372 | 0.932 |

## 4.2 Effect of oxygen evaluation on resulted reaction rates ratios

The oxygen cross sections significantly influence the macroscopic cross section of water at neutron energies above ~7MeV. Water thus defines the high-energy part of the neutron spectrum, which can influence the threshold reaction rates of the activation foils used.

To study this effect, the comparison between calculated and experimentally determined ratios of $^{58}$Ni(n,2n) and $^{19}$F(n,2n) to $^{27}$Al(n,α) has been performed. Calculations were performed using the same MCNP model and materials description in the ENDF/B-VIII library, except for $^{16}$O, which was used from different evaluations (ENDF/B-VII.1, JEFF-3.3, and JENDL-4). The comparison is listed for $^{58}$Ni(n,2n) to $^{27}$Al(n,α) in Table 14 and $^{19}$F(n,2n) to $^{27}$Al(n,α) in Table 15.

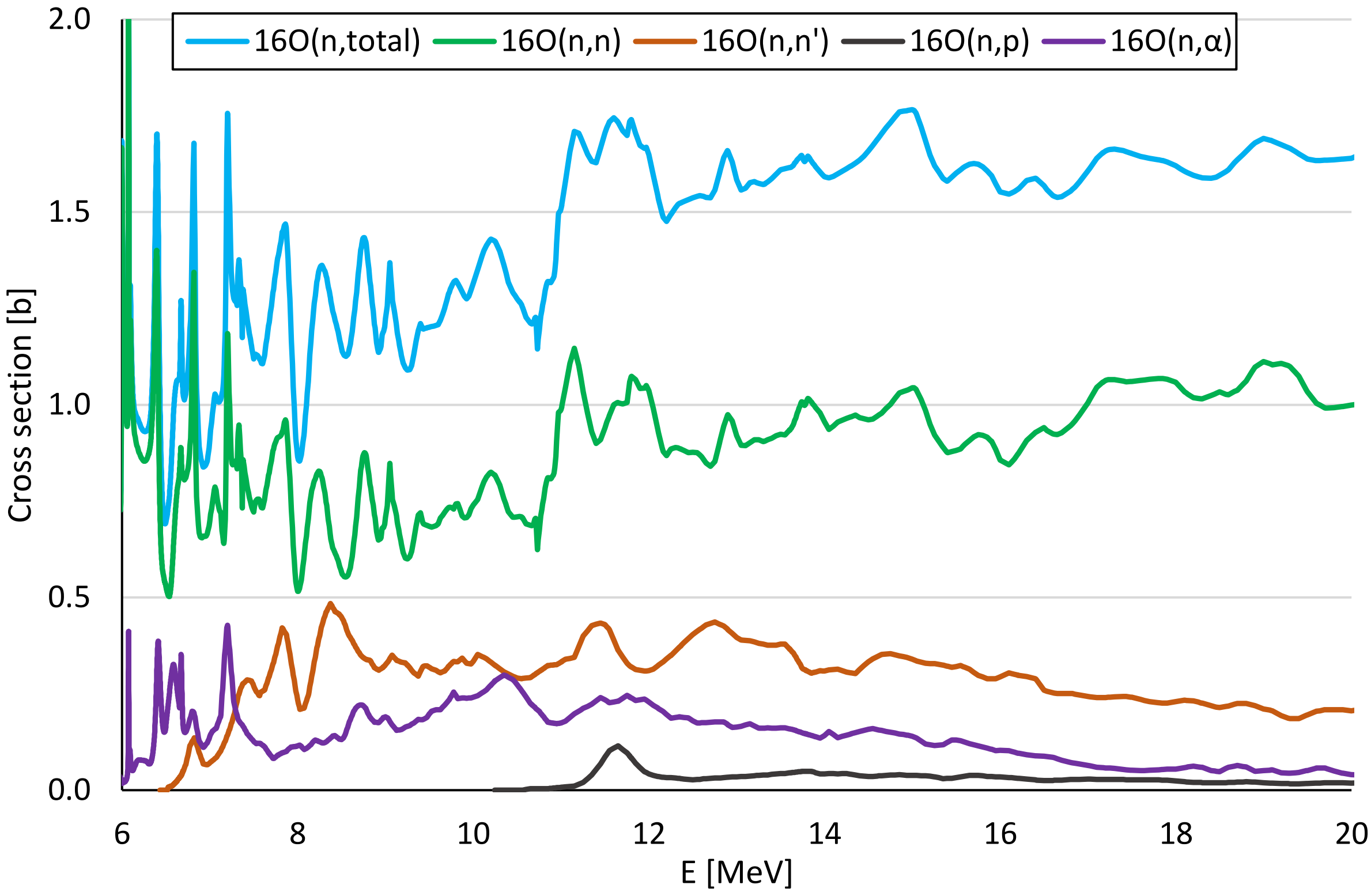

Figure 13: Comparison of cross sections of oxygen dominant reactions from ENDF/B-VIII in region above 6 MeV

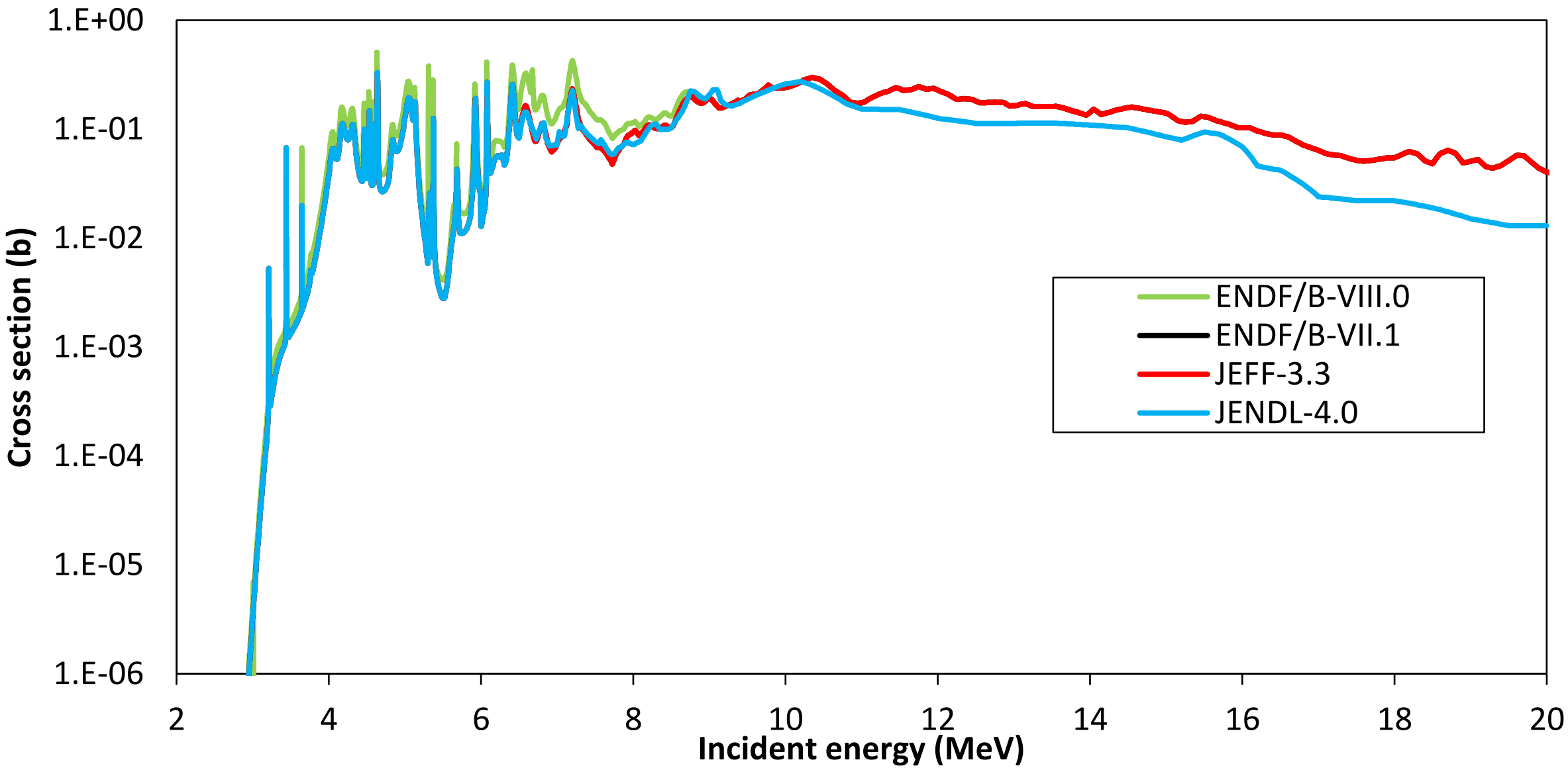

Figure 14: Comparison of $^{16}O(n, \alpha)$ in various evaluations

Table 14.: $^{58}Ni(n,2n)/^{27}Al(n,\alpha)$ reaction rates ratio using various oxygen evaluation

| | 1 | 2 | 3 | 4 | 5 | 6 |
|---|---|---|---|---|---|---|
| Experiment | 6.62E-3 | 5.70E-3 | 6.46E-3 | 5.79E-3 | 7.09E-3 | 8.70E-3 |
| ENDF/B-VIII | 6.75E-3 | 6.24E-3 | 6.58E-3 | 6.21E-3 | 7.53E-3 | 7.53E-3 |
| ENDF/B-VII.1 | 6.65E-3 | 6.24E-3 | 6.66E-3 | 6.25E-3 | | 7.97E-3 |
| JEFF-3.3 | 6.62E-3 | 6.22E-3 | 6.52E-3 | 6.35E-3 | 7.38E-3 | 7.91E-3 |
| JENDL-4 | 6.37E-3 | 6.36E-3 | 6.63E-3 | 6.21E-3 | 6.90E-3 | 7.91E-3 |
| Rel. unc. | 4.6% | 5.6% | 6.0% | 3.2% | 4.7% | 15.8% |

Table 15.: $^{19}$F(n,2n)/$^{27}$Al(n,α) reaction rates ratio using various oxygen evaluation

| | 1 | 2 | 3 | 4 | 5 | 6 |
|---|---|---|---|---|---|---|
| Experiment | 1.23E-2 | 1.13E-2 | 1.22E-2 | 1.14E-2 | 1.32E-2 | 1.44E-2 |
| ENDF/B-VIII | 1.31E-2 | 1.23E-2 | 1.29E-2 | 1.22E-2 | 1.43E-2 | 1.46E-2 |
| ENDF/B-VII.1 | 1.28E-2 | 1.22E-2 | 1.30E-2 | 1.23E-2 | x | 1.50E-2 |
| JEFF-3.3 | 1.29E-2 | 1.23E-2 | 1.26E-2 | 1.24E-2 | 1.40E-2 | 1.49E-2 |
| JENDL-4 | 1.24E-2 | 1.23E-2 | 1.28E-2 | 1.22E-2 | 1.33E-2 | 1.49E-2 |
| Rel. unc. | 3.4% | 2.3% | 3.2% | 2.4% | 5.4% | 2.7% |

It is worth noting that the $^{58}$Ni(n,2n) to $^{27}$Al(n,α) ratio in pure $^{235}$U($n_{th}$,f) PFNS is 5.82E-3. A very similar value was obtained by experiment in positions 2 and 4, which were fully surrounded by water. However, calculated values are consistently overpredicting the experiment by 6–12 %. At the boundary positions, experimental ratios are about 12% above the reference ratio in the PFNS, reflecting the fact that the neutron spectrum at the boundary positions is harder than $^{235}$U($n_{th}$,f) PFNS. In the $^{19}$F(n,2n) to $^{27}$Al(n,α) ratio, the trend is similar, and only the magnitude of difference from the PFNS ratio is lower than in $^{58}$Ni(n,2n) to $^{27}$Al(n,α) ratio as the threshold $^{19}$F(n,2n) is lower. Namely, the ratio in pure $^{235}$U($n_{th}$,f) PFNS is 1.16E-2, while in the core center is about 2% lower and at the boundary about 6% higher.
The comparison of experimental and evaluations results for oxygen data from different nuclear data libraries for the ratio $^{58}$Ni(n,2n) to $^{27}$Al(n,α) is shown in Table 16. In the core center, the calculation in all cases overpredicts the experiment; in the boundary positions, the agreement is better. Even in the position behind the water gap, the agreement is slightly better than in the central position. The comparison of $^{19}$F(n,2n) to $^{27}$Al(n,α) ratio is in Table 17. The trend is comparable to the previous case. Generally, the best agreement can be seen in JENDL-4 evaluation. However, the discrepancies in core center are observed in all evaluations.

Table 16.: C/E-1 of $^{58}$Ni(n,2n)/$^{27}$Al(n,α) reaction rates ratio

| | 1 | 2 | 3 | 4 | 5 | 6 |
|---|---|---|---|---|---|---|
| ENDF/B-VIII | 2.0% | 9.5% | 1.9% | 7.3% | 6.2% | -13.4% |
| ENDF/B-VII.1 | 0.4% | 9.4% | 3.1% | 8.0% | | -8.4% |
| JEFF-3.3 | 0.0% | 9.2% | 0.9% | 9.7% | 4.1% | -9.1% |
| JENDL-4 | -3.8% | 11.6% | 2.7% | 7.3% | -2.6% | -9.1% |
| Rel. unc. | 4.6% | 5.6% | 6.0% | 3.2% | 4.7% | 15.8% |

Table 17.: C/E-1 of $^{19}$F(n,2n)/$^{27}$Al(n,α) reaction rates ratio

| | 1 | 2 | 3 | 4 | 5 | 6 |
|---|---|---|---|---|---|---|
| ENDF/B-VIII | 6.5% | 8.8% | 5.7% | 7.0% | 8.3% | 1.4% |
| ENDF/B-VII.1 | 4.3% | 8.3% | 6.5% | 7.5% | | 4.2% |
| JEFF-3.3 | 4.5% | 8.7% | 3.5% | 8.8% | 5.9% | 3.6% |
| JENDL-4 | 0.5% | 8.9% | 5.0% | 7.0% | 1.0% | 3.6% |
| Rel. unc. | 3.4% | 2.3% | 3.2% | 2.4% | 5.4% | 2.7% |

## 5 Conclusions

In this work, a set of high-energy threshold dosimetry reactions were measured in central and boundary positions of the experimental reactor VR-1 core. Good agreement between calculation and experiment was obtained. This result confirms previous validation.

It was also shown that the neutron spectrum in the center of the VR-1 core formed by IRT-4M fuel in region 6–14 MeV is undistinguishable from $^{235}U(n_{th},f)$ PFNS. In the core boundary, the fast part of the spectrum above ~5MeVis significantly harder. In positions where there is water in between the fuel and the actual detector position, the spectrum hardening has even larger magnitude because the effective water thickness is larger. This result is confirmed by both calculation and experiment. Based on these results, it can be deduced that the fast part of neutron spectrum in compact cores formed by tubular type fuel is highly affected by actual position in the core and can eventually be harder than the $^{235}U(n_{th},f)$ prompt fission neutron spectrum.

The IRDFF-II neutron cross sections for 8 dosimetry threshold reactions averaged in the $^{235}U(n_{th},f)$ PFNS were compared against the measured SACS in 6 channels of the VR-1 reactor field. As a rule an agreement within 1-2 total uncertainties was observed, except for the reaction $^{197}Au(n,2n)$. For the latter the underestimation was found to increase up to 50% as foil location moves away from the reactor core center. It was found that the photo-nuclear reaction $^{197}Au(\gamma,n)$ contributes up to 11% to the production of $^{196}Au$ in comparison with $^{197}Au(n,2n)$ alone.

Further analysis and calculations have discovered that γ rays from thermal neutron capture on the steel components of VR-1 are the dominant source of gammas inducing photonuclear reactions of relevance. Finally, it was established that namely the 9 MeV γ-rays, i.e. the transition to the ground states in reaction $^{54}Fe(n_{th},\gamma_0)^{55}Fe$(g.s.), results to the additional production of $^{196}Au$. Validation of this γ-ray yield against the IAEA PGAA experimental database has however demonstrated deficiencies in the evaluated thermal capture gammas on iron of the ENDF/B-VIII.0 evaluation compared to ENDF/B-VII.1. These deficiencies result in a 40% underestimation of the $^{196}Au$ production.

Applying corrections both for the contribution of (γ,n) and the underestimation of $^{54}Fe(n_{th},\gamma_{gs})^{55}Fe$(g.s.) improved an agreement of calculations and measurements for the gold foil in the outcore irradiation channel within 38%, The remaining underestimation is attributed to the evaluated IAEA/PD-2019 $^{197}Au(\gamma,n)^{196}Au$ cross section in the first 2 MeV interval above the reaction threshold. This suggestion was confirmed by comparison with existing experimental data $^{197}Au(\gamma,n)^{196}Au$: in the energy interval of influence the IAEA/PD-2019 indeed underestimates measured data by about 30%. Such deficiency should be addressed by IAEA/PD-2019 evaluators.

## 6 Acknowledgements

The presented work was realized with use of reactors LVR-15 and LR-0, which were financially supported by the Ministry of Education, Youth and Sports - projects LM2018120, and the EU SANDA project funded under H2020-EURATOM-1.1 contract 847552. The authors would like to thank to VR-1 staff headed by F. Fejt for their effective help during the experiments and for the precise monitoring of the reactor power.

## 7 References

Abd 2010 A .El Abd, Measurement of the fission neutron spectrum averaged cross sections for the 95Mo(n, p)95Nb, 92Mo(n, α)89Zr, 90Zr(n, 2n)89Zr and 60Ni(n, p)60Co reactions Appl. Rad. and Isot., Vol. 68, Issue 10, (2010), pp. 2007-2012

Arribére, et al. 2001 M.A. Arribére, S. Ribeiro Guevara, P.M. Suárez, A.J. Kestelman, Threshold Reaction Cross Sections of Nickel Isotopes, Averaged Over a 235U Fission Neutron Spectrum", Nuclear Science and Engineering, 139, 2001, pp. 24–32

Bily et al., 2019 T. Bily, J. Rataj, O. Huml, O. Chvala, "Effect of kinetics parameters on transients calculations in external source driven subcritical VR-1 reactor", Ann. Nucl. Energy, 123 (2019), pp. 97-109

Boson et al 2008 J. Boson, G. Ågren, L. Johansson A detailed investigation of HPGe detector response for improved Monte Carlo efficiency calculations, Nuclear Instruments and Methods in Physics Research Section A, 587, 2008, pp 304-314
Brown et al 2018 D.A. Brown, M.B. Chadwick, R. Capote, et al., ENDF/B-VIII.0: the 8$^{th}$ major release of the nuclear reaction data library with CIELO-project cross sections, new standards and thermal scattering data, Nucl. Data Sheets, 148 (2018), pp. 1-142
Bruggeman et al. 1974 A. Bruggeman W. Maenhaut, and J. Hoste, "The Total Average Cross Section for the Reactions 58Ni(n,np)57Co, 58Ni(n,pn)57Co, and 58Ni(n,d)57Co in a Fission Type Reactor Spectrum," Radiochem. Radioanal. Lett., 18, 87 (1974).
Czakoj 2021 Measurement of prompt gamma field above the VR-1 water level, T. Czakoj, M. Košťál, Z. Matěj, E. Losa, J. Šimon, F. Mravec, F. Cvachovec; EPJ Web Conf., 253 (2021) 04014, https://doi.org/10.1051/epjconf/202125304014
Dorval et al 2006 E. L. Dorval, M. A. Arribére, S. Ribeiro Guevara, I. M. Cohen, A. J. Kestelman, R. A. Ohaco, M. S. Segovia, A. N. Yunes, M. Arrondo, Fission neutron spectrum averaged cross sections for threshold reactions on arsenic, Journal of Radioanalytical and Nuclear Chemistry, Vol. 270, No.3 (2006), pp. 603–608
Dryak et al 2006 P. Dryak, P. Kovar, Experimental and MC determination of HPGe detector efficiency in the 40–2754 keV energy range for measuring point source geometry with the source-to-detector distance of 25 cm, Appl. Rad. and Isot., 64, 2006, pp 1346-1349
Firestone et al. 2017 R.B. Firestone, T. Belgya, M. Krticka et al., Thermal neutron capture cross section for $^{56}$Fe(n,$\gamma$), Phys. Rev. C95 (2017) 014328
Franke et al. 2001 S.C. Frankle, R.C. Reedy, P.G. Young, Improved Photon-Production Data for Thermal Neutron Capture in the ENDF/B-VI Evaluations, Report LA-13812, Los Alamos 2001
Frybort 2020 J. Frýbort, P. Suk, F. Fejt, Designing Stainless Steel Reflector at VR-1 Training Reactor, EPJ Web Conf. 239 (2020) 17009. https://doi.org/10.1051/EPJCONF/202023917009.
T. Goorley et al 2012 T. Goorley, et al., "Initial MCNP6 Release Overview", Nuclear Technology, **180**, pp 298-315 (Dec 2012)
Hara et al. 2007 K.Y. Hara, H. Harada, F. Kitatani et al., "Measurements of the 152Srn($\gamma$,n) Cross Section with Laser-Compton Scattering $\gamma$ Rays and the Photon Difference Method", Jour. of Nucl. Sci. and Techn. 44 (2007) 938
Huml et al., 2013 O. Huml, J. Rataj, T. Bily, "Application of MCNP for neutronic calculations at VR-1 training reactor Proceedings of SNA + MC 2013", Joint International Conference on Supercomputing in Nuclear Applications + Monte Carlo, Paris (2013), 10.1051/snamc/201405103
Itoh et al. 2011 O. Itoh, H. Utsunomiya, H. Akimune et al., "Photoneutron cross sections for Au revisited: measurements with laser compton scattering gamma-rays and data reduction by a least-squares method", Jour. of Nucl. Sci. and Techn. 48 (2011) 834
Kawano et al. 2020, T. Kawano, Y. S. Cho, P. Dimitriou, et al., "IAEA Photonuclear Data Library 2019", Nuclear Data Sheets 163 (2020) 109 - 162.
Kostal et al 2018 M. Košťál, Z. Matěj, E. Losa, On similarity of various reactor spectra and $^{235}$U prompt fission neutron spectrum, Appl. Rad. and Isot., Vol. 135 (2018), pp. 83–91
Kostal et al 2018b M. Košťál, M. Schulc, J. Šimon et al., Measurement of various monitors reaction rate in a special core at LR-0 reactor, Ann. of Nucl. En., Vol. 112, (2018), pp. 759–768
Kostal et al 2020 M. Košťál, M. Schulc, E. Losa et al., A reference neutron field for measurement of spectrum averaged cross sections, Ann. of Nucl. Energy, Vol. 140, (2020), 107119

Kostal et al 2021 M. Kostal, E. Losa, M. Schulc, J. Simon, T. Bily, V. Rypar, M. Marecek, J. Uhlír, T. Czakoj, R. Capote, A. Trkov, S. Simakov, Validation of IRDFF-II library in VR-1 reactor field using thin targets, Ann. of Nucl. Energy 158, (2021), 108268
Losa et al. 2021 Losa, E., Košťál, M., Stefanik, M. et al., Validation of the fast neutron field in the radial channel of the VR-1 reactor, Journal of Nuclear Engineering and Radiation Science, 2021, 7(2), 021503-1
Maidana et al 1994 N. L. Maidana, M. S. Dias, L. P. Geraldo, Measurements of U-235 Fission Neutron Spectrum Averaged Cross Sections for Threshold Reactions, Radiochimica Acta, 64, 7 - 9 (1994)
Mannhart 2002 W. Mannhart, Validation of Differential Cross Sections with Integral Data , Report INDC(NDS)-435, pp.59-64, IAEA, Vienna, September 2002
Mughabghab 2018 S.F. Mughabghab, Atlas of Resonances; 6th edition, Elsevier Science, 2018
OECD Nuclear Energy Agency, "Collaborative International Evaluated Library Organisation (CIELO) Pilot Project", WPEC Subgroup 40 (SG40) (see www.oecdnea.org/science/wpec/sg40-cielo/)
PGAA 2007 "Database for Prompt Gamma-ray Neutron Activation Analysis" (PGAA), IAEA Report STI/PUB/1263, Vienna 2007.
Rataj et al., 2014 J. Rataj, O. Huml, L. Heraltová, T. Bílý, "Benchmark experiments for validation of reaction rates determination in reactor dosimetry", Radiat. Phys. Chem., 104 (2014), pp. 363-367
Steinnes 1970 E. Steinnes, Cross-sections of some (n,2n) reactions induced by reactor fast neutrons, Radiochimica Acta 13, 1970, pp. 169-171
Stetcu et al., 2020, I. Stetcu, M. B. Chadwick, T. Kawano, P. Talou, R. Capote, "Evaluation of the Prompt Fission Gamma Properties for Neutron Induced Fission of 235,238U and 239Pu", Nuclear Data Sheets 163 (2020) 261–279
Suarez et al 1997 P. M. Suárez, M. A. Arribére, S. Ribeiro Guevara, A. J. Kestelman, Experimental and Calculated Scandium Threshold Reaction Cross Sections Averaged over a 235U Fission Neutron Spectrum, Nuclear Science and Engineering, Vol. 127, pp. 245–261, (1997)
Tomarchio et al 2009 E. Tomarchio, S.Rizzo, Coincidence-summing correction equations in gamma-ray spectrometry with p-type HPGe detectors, Radiation Physics and Chemistry, Vol. 80 (2011), pp. 318-323
Trkov et al 2001 A. Trkov, "Program RR_UNC - calculates uncertainties in reaction rates and cross sections"; code available on https://www-nds.iaea.org/IRDFF/
Trkov et al 2020 A. Trkov, P. J. Griffin, S. P. Simakov et al., IRDFF-II: A New Neutron Metrology Library, Nuclear Data Sheets, Vol. 163, 2020, pp. 1–108, data available on https://www-nds.iaea.org/IRDFF/.
Werner et al 2017 C. J. Werner, J. S. Bull, C. J. Solomon et al., "MCNP Users Manual - Code Version 6.2", LA-UR-17-29981 (2017).
WÖLFLE et al 1980 R. Wölfle and S. M. Qaim, "Measurement of Fission Neutron Spectrum Averaged Cross-Sections for Some (n, p), (n, n'p) and (n, a) Reactions on Nickel and Chromium," Radiochim. Acta, 27, 65 (1980).
ZAIDI et al. 1993 J. H. Zaidi, H. M. A. Karim, M. Arif, I. H. Qureshi, and S. M. Qaim, "Measurement of Fission Neutron Spectrum Averaged Cross-Sections on Nickel: Small Scale Production of 57Co in a Nuclear Reactor," Radiochim. Acta, 60, 169 (1993)